\documentclass[useAMS,usenatbib]{mn2e}
\usepackage{graphicx,amsmath,multirow,amssymb}
\usepackage{subfig}
\usepackage{natbib}
\newcommand{\comment}[1]{}

\def\simgt{\lower.5ex\hbox{$\; \buildrel > \over \F sim \;$}}
\def\simlt{\lower.5ex\hbox{$\; \buildrel < \over \sim \;$}}

\title[Evolved stars in IC10]{Evolved stars in the Local Group galaxies - II. 
AGB, RSG stars and dust production in IC10}
\author[Dell'Agli et al.]{F. Dell'Agli$^{1,2}$, M. Di Criscienzo$^3$, P. Ventura$^{3}$, M. Limongi$^{3}$,
\newauthor
 D. A. Garc\'{\i}a--Hern\'andez$^{1,2}$, E. Marini$^{3,4}$, C. Rossi$^{5}$\\
$^{1}$Instituto de Astrof\'{\i}sica de Canarias (IAC), E-38200 La Laguna, Tenerife, Spain \\
$^{2}$Departamento de Astrof\'{\i}sica, Universidad de La Laguna (ULL), E-38206 La Laguna, Tenerife, Spain \\
$^3$INAF -- Osservatorio Astronomico di Roma, Via Frascati 33, 00040, Monte Porzio Catone (RM), Italy \\
$^4$Dipartimento di Matematica e Fisica, Universita degli Studi ``Roma Tre", Via della Vasca Navale 84, I-00146 Roma, Italy\\
$^5$Dipartimento di Fisica, Universit\`a di Roma ``La Sapienza'', P.le Aldo Moro 5, 00143, Roma, Italy \\
}

\begin{document}

\date{Accepted, Received; in original form }

\pagerange{\pageref{firstpage}--\pageref{lastpage}} \pubyear{2012}

\maketitle

\label{firstpage}

\begin{abstract}
We study the evolved stellar population of the Local Group galaxy IC10, with the aim of characterizing the
individual sources observed and to derive global information on the galaxy, primarily the
star formation history and the dust production rate. To this aim, we use evolutionary sequences 
of low- and intermediate-mass ($M < 8~M_{\odot}$) stars, evolved through the asymptotic 
giant branch phase, with the inclusion of the description of dust formation. We also use
models of higher mass stars. 
From the analysis of the distribution of stars in the observational planes obtained with 
IR bands, we find that the reddening and distance of IC10 are $E(B-V)=1.85$ mag and 
$d=0.77$ Mpc, respectively.
The evolved stellar population is dominated by carbon stars, that account for $40\%$ of the
sources brighter than the tip of the red giant branch. Most of these stars descend from
$\sim 1.1-1.3~M_{\odot}$ progenitors, formed during the major epoch of star formation,
which occurred $\sim 2.5$ Gyr ago.  
The presence of a significant number of bright stars indicates that IC10 has been site of
significant star formation in recent epochs and currently hosts a group of massive stars
in the core helium-burning phase. 
Dust production in this galaxy is largely dominated by carbon stars; the overall dust
production rate estimated is $7\times 10^{-6}~M_{\odot}$/yr. 
\end{abstract}


\section{Introduction}
Dwarfs galaxies are the dominant component among the structures populating the Local
Group (LG). They are characterized by a significant diversity in terms of structural properties,
star formation history (SFH) and metal enrichment (see e.g. Weisz 2014). The understanding of 
the origin of these heterogeneities proves crucial for a number of astrophysical contexts, 
ranging from pure stellar evolution grounds to some fundamental aspects regarding the 
formation and the evolution of galaxies.

Among the dwarfs galaxies in the Local Group IC10 is one of the most interesting systems,
given its large total mass ($\log(M/M_{\odot}) \sim 8.5$; Vaduvescu et al. 2007), 
which makes it one of the most massive and luminous dwarfs. Furthermore, the large 
H$\alpha$ luminosity \citep[e.g.][]{mateo98,kennicutt08,tehrani17} and the very large 
density of Wolf-Rayet stars \citep{massey92, massey95,massey02}, a factor $\sim 20$ higher 
than in the Large Magellanic Cloud (hereafter LMC), suggest that IC10 is currently 
experiencing vigorous star formation.

\citet{sanna08a, sanna09} used Hubble Space Telescope (HST) data to derive the distance of 
IC10 (distance modulus $\mu=24.60\pm0.15$ mag), 
based on the magnitude of the tip of the red giant branch (hereafter TRGB); the authors
also investigated the age and metallicity distribution, by comparing ACS and WFPC2 data
with results from evolutionary calculations of main sequence, red giant and horizontal
branch stars. A complementary approach to reconstruct the history of the star formation 
in IC10 was followed by \citet{magrini09}, based on its planetary nebulae (PNe) and
HII regions populations. The first uniform derivation of 40 LG dwarf galaxies SFHs was 
recently presented by \citet{weisz14}, based on analysis of colour-magnitude diagrams 
constructed from archival HST/WFPC2 imaging. IC 10 was part of this study, where the main 
peak of the SFH is found around 1.5-4 Gyr, followed by a significant star formation rate 
(SFR) even in the most recent epoch ($\sim$400Myr and 10 Myr).

All these results clearly evidence the suitability of IC10 as excellent target to study 
the evolved population in dwarf galaxies. A detailed analysis is now possible thanks to 
the significant improvements in the modelling of the asymptotic giant branch (AGB) phase 
of stars of low- and intermediate-mass. Indeed, the description of the stellar
structure has been completed in recent times with the modelling of the dust formation 
in the circumstellar envelopes \citep{ventura12a, ventura12b, ventura14, nanni13, nanni14}, 
a key ingredient to interpret IR data of cool stars.

The comparison between IR data and AGB models which
consider dust formation proved useful in several studies aimed at understanding the
properties of dust-enshrouded objects in the Magellanic Clouds (MCs)
(e.g. Boyer et al, 2015c, Kraemer et al. 2017).
Recent studies used star+dust systems to characterise the 
entire AGB population of the MCs, in terms of mass, age and metallicity distribution 
\citep{flavia14, flavia15a, flavia15b, nanni17}. These studies turned out to be extremely useful 
not only to interpret the observed stellar population, but also to infer important
information regarding the efficiency of some chemical phenomena driving the evolution
through the AGB phase, the modality with which dust formation occurs and the absorption and
scattering properties of the most relevant dust species formed in the AGB winds
\citep{ventura15, ventura16, nanni16}.

While the MCs have so far been proved to be the best laboratories to study the AGB phase,
owing to their low reddening and known distances, the increasing availability of 
high-quality data in other LG galaxies will soon allow a better understanding of the main 
properties of AGB stars.
An exhaustive analysis of the evolution of this class of objects demands that further
environments external to the MCs are explored; this is even more important given
the forthcoming launch of the James Webb Space Telescope (\emph{JWST}), which will make available 
data of thousands of AGB stars from several galaxies in the Local Group.
The opportunities offered by the \emph{JWST} mission for these studies is
discussed in \citet{jones17}.

Early attempts to use IR data to derive information on the evolved population of 
environments beyond the MCs were done, e.g., for M33 \citep{javadi17}, NGC 147 and
NGC 185 \citep{hamedani17}. In a study focused on the evolved stellar population
of IC 1613, \citet{flavia16} interpreted the AGB stars of this galaxy, providing a 
characterization of the sources observed, in terms of mass, chemical composition and 
formation epoch of the progenitors.

In this work we apply the same approach used in \citet{flavia16} to study the evolved 
population of IC10. This is possible thanks to the combined availability of near-infrared 
(NIR) data published by \citet{gerbrandt15}
and of \emph{Spitzer} mid-infrared data by the survey of Dust in Nearby Galaxies with
\emph{Spitzer} (DUSTINGS; Boyer 
et al. 2015a,b). We carry out a statistical analysis of the distribution 
of stars in the colour-colour and colour-magnitude planes (obtained with various 
combinations of NIR and mid-IR \emph{Spitzer} filters) to refine the SFH, allowing the best 
compatibility between models and observations. 
We show how this methodology also allows the determination of the reddening and the 
distance of the galaxy. To our knowledge, in the context of the interpretation of evolved 
and dusty populations, this is the first work that considers the possible 
contamination from RSG stars in the regions of the observational planes populated by 
AGB stars. This issue is particularly relevant in a starburst galaxy such as IC10.

The paper is organized as follows: Section \ref{sample} is devoted to present the observational 
sample considered; the stellar evolution and dust formation models
are briefly described in Section \ref{ATON}; the main properties of the AGB stars used in the
present work are discussed in Section \ref{models}. In Section \ref{calib} we describe the 
construction of the synthetic population, the comparison of
the observed and the expected distribution of the evolved stars in different colour-colour
and colour-magnitude planes, in order to find the reddening, distance and to 
reconstruct the SFH of the galaxy. The characterization of the evolved stellar 
population of IC10 is addressed in Section 6, while the most obscured sources are described
in Section 7. Section 8 is dedicated to describe the dust produced by AGB stars in IC10
and to estimate the overall dust production rate of the galaxy; the PNe population of IC10,
specifically the objects for which both the nitrogen and oxygen abundances have been
determined, is interpreted in Section 9. Finally, the conclusions are given in Section 10.

\section{Observational samples}
\label{sample}
To perform a quantitative analysis of the evolved stellar population of IC10 we rely
on the most complete samples of stars available in the literature. We focus our attention on 
the NIR and mid-IR wavelength range, where the emission for the majority of these stars peaks. 

\citet{gerbrandt15} provided {\it J}, {\it H} and {\it K} data for a wide sample of stars 
in IC10 within an area of 0.75\,deg$^2$, using the Wide Field Camera on the 3.8m United 
Kingdom Infrared Telescope, during a single observing run. Moreover, the DUSTiNGS survey 
\citep{boyer15a} was specifically designed to identify dust-producing AGB stars, 
made available mid-IR magnitudes for hundreds of stars in dwarf galaxies in the Local 
Group. Among the others, IC10 has been observed with the InfraRed Array Camera in the 
[3.6] and [4.6] filters during two epochs, to reduce the effects of variability. 

The wide opportunities offered by the simultaneous analysis of NIR 
and mid-IR data to study evolved stars and to distinguish different classes 
of objects was explored in a series of works \citep{boyer11, mcdonald12, woods11}.
This approach was also used by \citet{flavia16} to understand the distribution 
and extension of the evolved stellar population of IC1613 in the various observational 
planes. In particular, in the 
colour-magnitude ($K-[4.5]$, [4.5]) plane, carbon and oxygen-rich stars occupy well defined 
and distinct regions. An important outcome of the study by \citet{flavia16} is that
the analysis of the distribution of stars in this plane allows  
to infer useful constrains on the SFH of the galaxy during the last Gyrs and the characterization 
of the dust produced by the individual sources.  

In order to carry out a similar analysis in IC10, we first homogenized the two 
astrometries, by applying a corrective factor based on the brightest stars of both 
catalogues; we then cross-correlated  the \citet{gerbrandt15} and \citet{boyer15a} samples, 
using a matching radius of 1.2 arcsec, which is the largest spatial resolution of Spitzer; 
when multiple matches occurred, only the closest match with the brigthest star was kept.

From the analysis 
described in Sections \ref{reddening} and \ref{distance}, we found and adopt a distance of 
0.77\,Mpc and reddening $E(B-V)=1.14$ mag, values in the range of the latest estimates from the 
literature \citep[e.g.][]{richer01,demers04,sanna08a, sanna08b}. 

\begin{figure}
\begin{minipage}{0.48\textwidth}
\resizebox{1.\hsize}{!}{\includegraphics{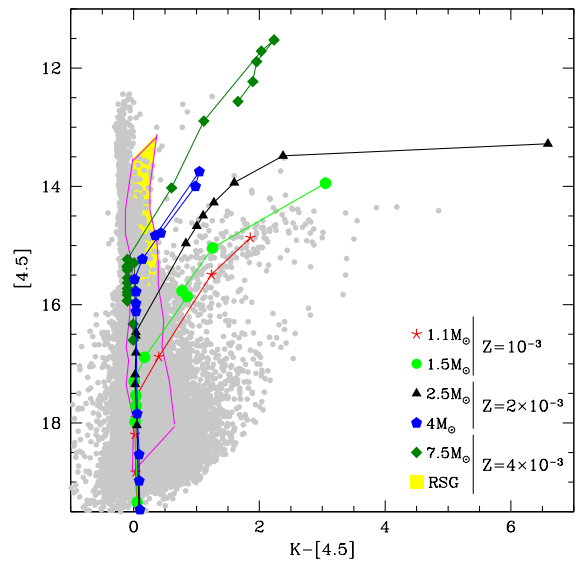}}
\end{minipage}
\vskip-50pt
\caption{The evolutionary tracks of stars of various mass and metallicity are overimposed
to the observations in the colour-magnitude $(K-[4.5],[4.5])$ plane. The yellow 
region, delimited by a magenta track, points the locus of the plane populated by
stars of mass in the range $6-20~M_{\odot}$ during the core helium-burning phase.
See the text for the choices regarding the assumed reddening and distance of IC10.}
\label{ftracks}
\end{figure}

The final sample is composed
of 20399 sources. Fig.~\ref{ftracks} shows the distribution of the stars in this sample,
indicated as grey points, in the $(K-[4.5],[4.5])$ colour-magnitude plane, to which we will 
refer hereafter as CMD. In order to clean the sample from the majority of the foreground 
objects, in the following analysis we will consider only stars with $(J-H)_0 >0.7$ mag, 
which, according to
\citet{gerbrandt15}, removes about $96\%$ of all the foreground contaminants. 
The removal of a small fraction of genuine IC10 stars by this cut makes no difference 
to the subsequent analysis. We will
also exclude all the stars with $K_0>$18.3 mag, to limit the
present analysis to the stars brighter than the TRGB. With these assumptions we obtain a
final sample composed of $\sim$3680 objects. A detailed comparison bewteen our theoretical 
expectations and the observations in this plane is presented in Section \ref{sint}.

To prevent the possibility that we miss a large fraction of the objects with the 
highest degree of obscuration, which we expect to have extremely faint {\it K} magnitudes, we 
take into account the stars identified as "extreme" (xAGB) by \citet{boyer15b}. These 
sources, included in the \emph{Spitzer} sample previously described, were classified on the basis 
of their variability and of their red ($[3.6]-[4.5]>0.1$) colours. Among all the galaxies 
investigated within DUSTiNGS, IC10 harbors the largest population of extreme stars. The chemical 
composition of these stars is unknown, thus their study is important to deduce the main 
properties of the progenitors. Furthermore, considering their large IR emission, suggesting 
significant dust production in their circumstellar envelopes, the characterization of these 
objects proves crucial to determine the overall dust production rate in IC10. While 
this choice allows us to provide a robust evaluation of the global dust production rate,
we believe important here to underline that the present analysis might likely provide
a lower limit of this quantity, as a paucity of extremely obscured stars, similar to those 
observed in the LMC by \citet{gruendl08}, with very low fluxes in the spectral region around
$4.5 \mu$m, might be not present in the \citet{boyer15b} sample.

The study by \citet{boyer15b} has been recently completed by \citet{boyer17}, who 
presented a follow-up survey with WFC3/IR on the Hubble Space Telescope (HST), and 
identified a total of $\sim 26$ dusty M-type stars, most of which belong to IC10.
These stars, characterized by a large IR excess, with IR colours up to
$[3.6]-[4.5] \sim 1$, are mostly confined in the inner regions of the galaxy, 
suggesting that they are more massive than the dominant C-rich obscured population.

Finally, \citet{lebo12} characterized stars in IC10 by their silicate dust 
features, to identify O-rich stars and to distinguish between AGB
and RSG stars. The detection of the silicate dust features was based on the
mid-IR spectra obtained with the IR Spectrograph (IRS) onboard of \emph{Spitzer}.

The discussion of all these samples on the basis of our synthetic population is presented in 
Sect. \ref{dustystars}.
Before addressing the characterization of the individual sources observed, we briefly present 
the evolutionary models adopted for the analysis and we discuss the 
reddening, distance and SFH of IC10, which allow the best agreement between the observational 
evidence and the theoretical framework.

\section{Modelling the AGB phase with dust formation and RSG stars}
\label{ATON}
To investigate the evolved stellar population of IC10 we consider models of stars with
initial masses in the range $0.8-8~M_{\odot}$ and metallicity\footnote{We indicate with
$Z$ the mass fraction of all the species but hydrogen and helium present initially in the 
star, thus characterizing the mixture of the gas from which the stars formed. Note that 
the overal metallicity increases during the AGB phase of the stars exposed to carbon 
enrichment in the surface layers} $Z=1,2,4 \times 10^{-3}$.
The upper limit in mass is due to the ignition of carbon burning in stars of mass above 
$8~M_{\odot}$; these stars will finally undergo core collapse, without evolving
through the AGB phase.

The models used here are presented and discussed in \citet{flavia16}, where all the
details of the AGB evolution and dust formation modelling are given. The description of the
AGB phase is essentially made up of three different steps: a) the modelling of the AGB
evolution, via the code {\sc ATON} \citep{ventura98}; b) we describe the dust formation process 
for some evolutionary phases chosen along the AGB, by means of the schematization by the 
Heidelberg group \citep{fg06}, which was applied by \citet{ventura12a, ventura12b, ventura14};
c) the infrared colours are obtained based on the values of the physical quantities
of the star and on the dust present in the envelope, by means of the DUSTY code
\citep{dusty}.

We believe important to stress here that to undertake this kind of analysis, based on the
interpretation of observations in the near- and mid-IR spectral region, the
description of the dust formation process is mandatory, because the infrared emission
by the star is determined by the number density and the grain size distribution of the dust
particles formed in the circumstellar envelope.

This having said, we remind that the present generation of AGB+dust models must be
considered as a preliminary step towards a more physically sound description, 
considering that the description proposed by
\citet{fg06} is based on a grey treatment of extinction and neglects the effects of pulsation,
which might affects the results obtained significantly \citep{bladh12, bladh13, bladh15, 
bladh17}. More important, the rate of mass loss in these models is assumed apriori, thus
it is not a consequence of the dust formation process; this makes the results obtained
critically dependent of the description of mass loss, which affects significantly both
the main properties of AGB evolution \citep{karakas14} and the formation of dust 
\citep{ventura14}. The choice of the optical constants also play an important role,
as shown, e.g. in \citet{nanni16}. Finally, the present description assumes that the
wind expands isotropically from the surface of single stars, thus neglecting the effects 
of the presence of a companion, confirmed by recent analysis of resolved AGB stars
\citep{ramstedt14, richards14, rau15}.

Consistently with the previous works on this argument, we will quantify the degree
of obscuration of the stars by the optical depth at $10 \mu$m, $\tau_{10}$, which is
determined by the radial distribution of the dust particles in the surroundings of the star
(see Section 2.3 in Dell'Agli et al., 2015a).

As stated in the introduction, because IC10 is a starbust galaxy, we considered 
the presence of RSGs in the sample observed. Stars with masses 
$\rm M\sim 6-20~\rm M_{\odot}$ evolve on time scales shorter than 70 Myr and during the 
helium-burning phase evolve at luminosities in the range 
$3\times 10^3 - 1.4\times 10^5$L$_{\odot}$. During the red part of the loop characterizing 
the core helium burning phase, the position of these objects in the colour-magnitude diagram 
partly overlaps with AGB stars. To give an estimate of the contamination from
RSG stars we considered the core helium-burning phase of the models mentioned earlier in 
this section, presented in \citet{flavia16}, with mass in the range 6-8 M$_{\odot}$. The 
RSG evolution of stars with masses $M \geq 9~M_{\odot}$ was described 
by means of the rotating version of the FRANEC code (the interested reader is referred to 
\citet{chieffi13} for more details). 

Given the uncertainties related to the mass loss and 
the dust production mechanisms in the winds of this class of objects, it is not 
possible to clearly model the distribution of these stars in the colour-magnitude diagrams. 
We tentatively assumed a variety of degrees of obscuration, described by optical depths in 
the range $0 < \rm \tau_{10} < 0.2$. Recently, \citet{groenewegen18} estimate $\rm \tau_{0.5}$ 
for $\sim 20$ supergiant stars in the SMC, fitting the SEDs and IRS spectra with a dust 
radiative-transfer model. Their results are in agreement with the degrees of obscuration adopted 
in this work. This adds more robustness to our assumption, considering the similarity in 
the metallicity of distribution of stars in IC10 and in the SMC. While the presence of 
heavily obscured RSG stars cannot be completely ruled out 
\citep[e.g.][]{jones15,goldman17,groenewegen18}, this conservative assumption on the range of
$\rm \tau_{10}$ values reached by core helium-burning stars allow us to identify a region 
in the CMD (indicated with a yellow shading delimited by magenta lines in Fig. \ref{ftracks}), 
where we expect that most (if not all) of these stars are found.

\section{AGB stars: evolution properties and IR colours}
\label{models}
The most recent and complete study of the SFH of IC10 is provided by \citet{weisz14}; 
we consider their results as a starting point to compute the synthetic population of 
evolved stars. According to \citet{weisz14} the main peak of the SFH occured around 1.5-4 
Gyr ago (see Fig.\ref{fsfh}), when low-mass stars of metallicity $Z \sim 10^{-3}$ formed. 
These are the progeny of most of the AGB population currently evolving in IC10. We describe 
here their main properties, as they evolve through the AGB phase, to ease the interpretation 
of the observed distribution of stars in the colour-colour and colour-magnitude planes.

Before entering the discussion of the evolutionary properties, we believe important
to stress that the description of the AGB phase of these stars, of initial mass below 
$\sim 3~M_{\odot }$, is fairly robust. This is the result, which holds for various metallicities, 
shown in a series of recent papers \citep{ventura16, ventura18}.
Conversely, the modelling of their higher mass counterparts is more uncertain, as it is
heavily affected by the modelling of the convective instability \citep{ventura15, ventura18}.

\begin{figure*}
\begin{minipage}{0.48\textwidth}
\resizebox{1.\hsize}{!}{\includegraphics{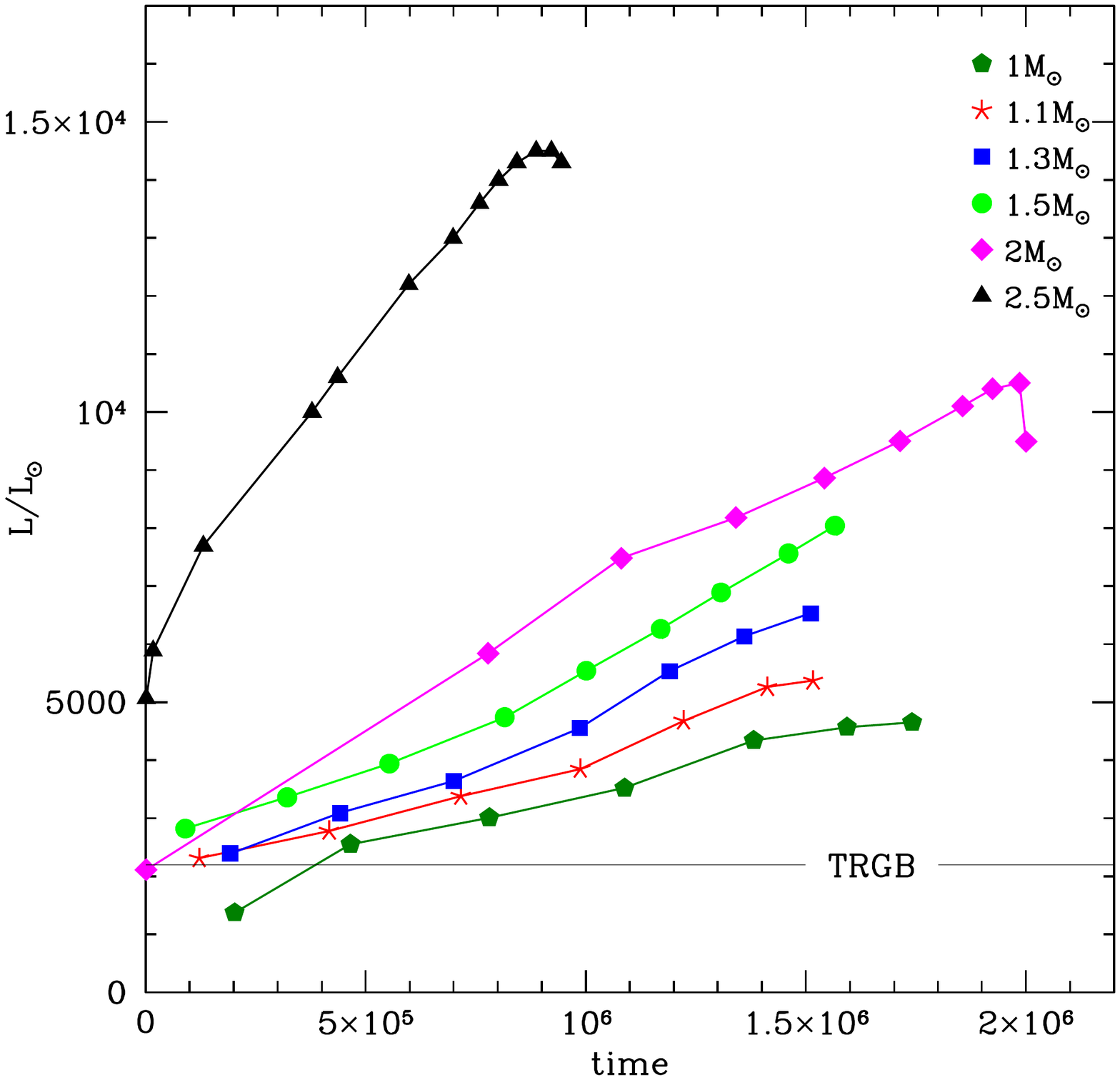}}
\end{minipage}
\begin{minipage}{0.48\textwidth}
\resizebox{1.\hsize}{!}{\includegraphics{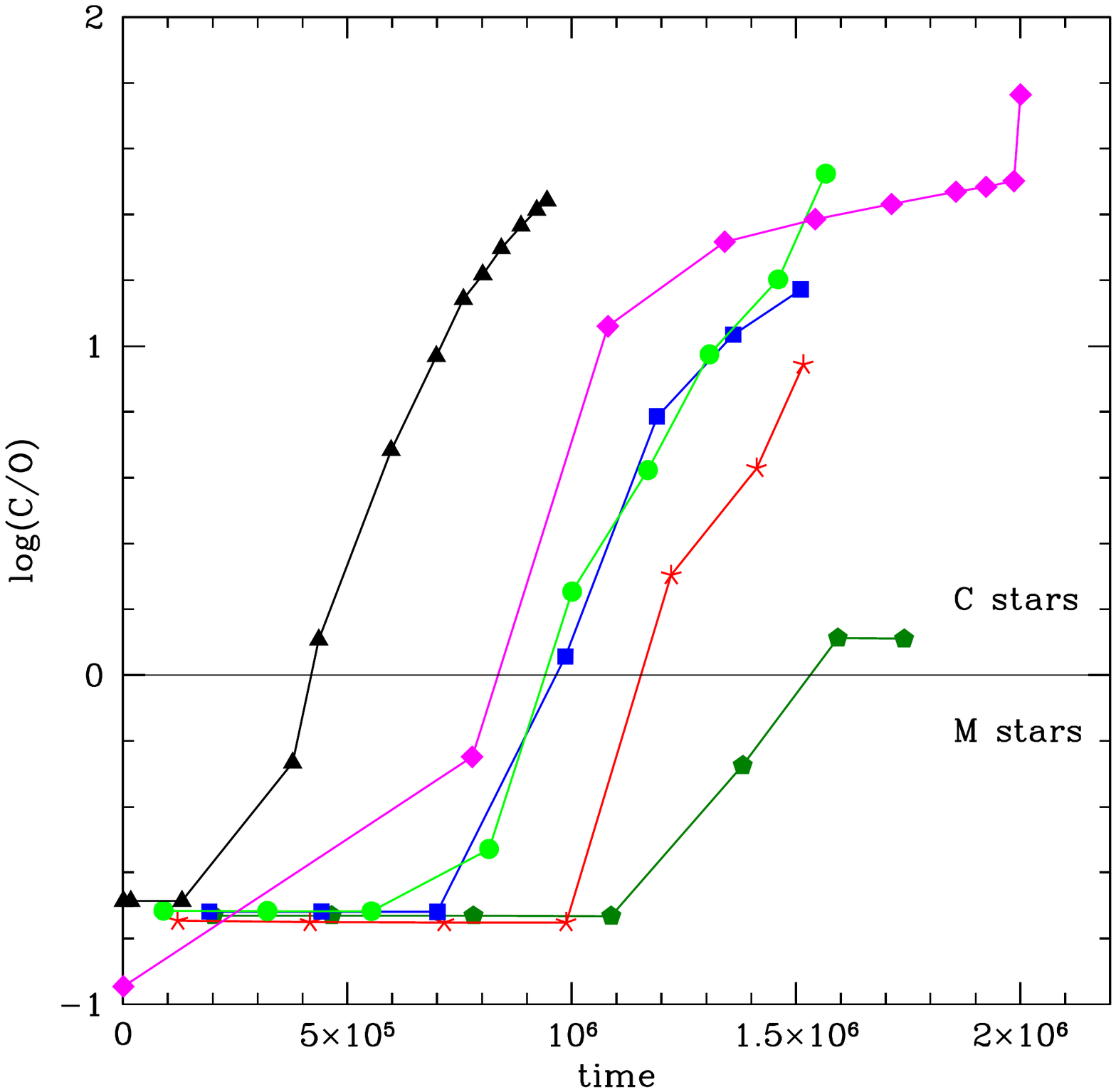}}
\end{minipage}
\vskip-70pt
\begin{minipage}{0.48\textwidth}
\resizebox{1.\hsize}{!}{\includegraphics{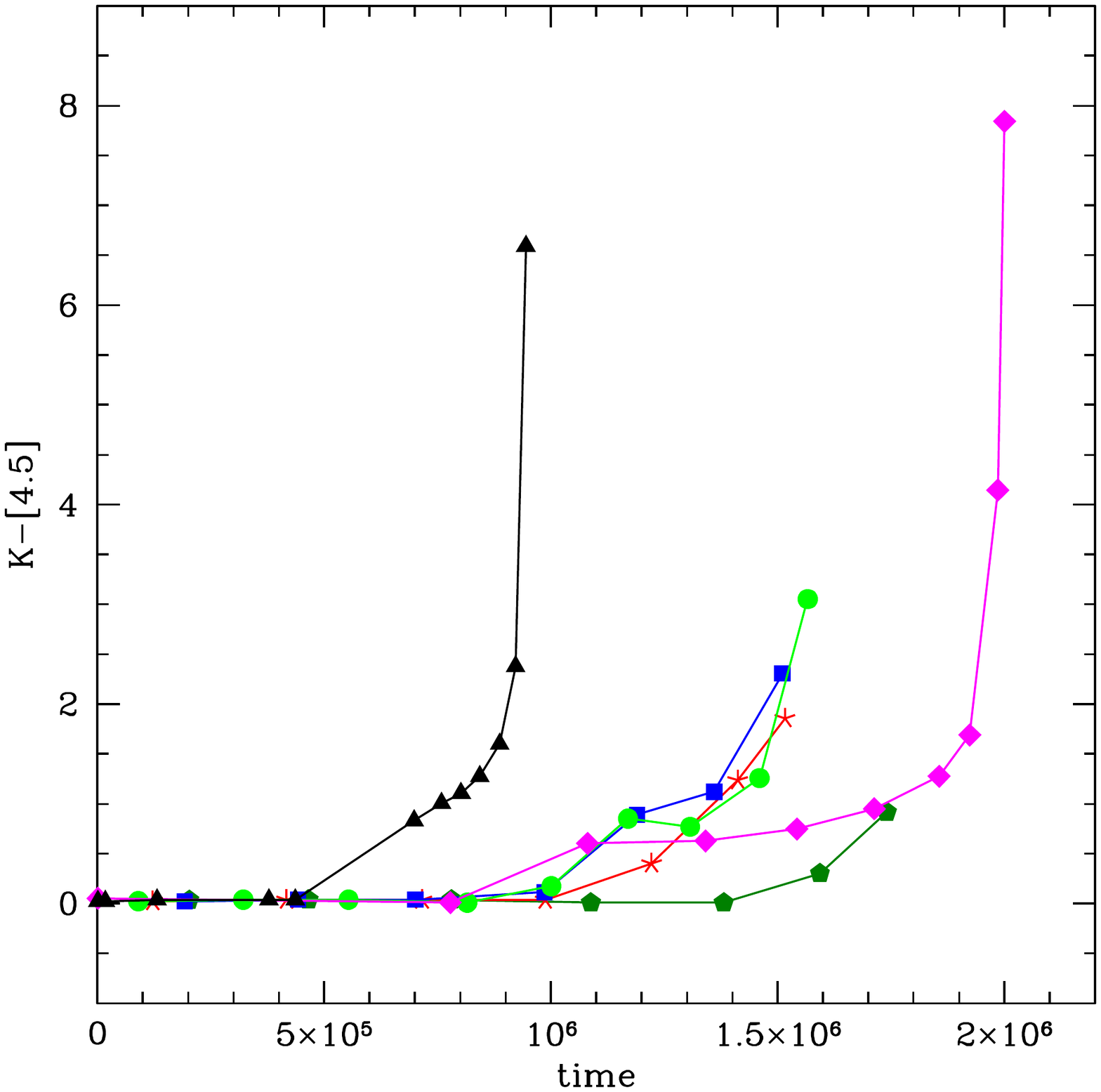}}
\end{minipage}
\begin{minipage}{0.48\textwidth}
\resizebox{1.\hsize}{!}{\includegraphics{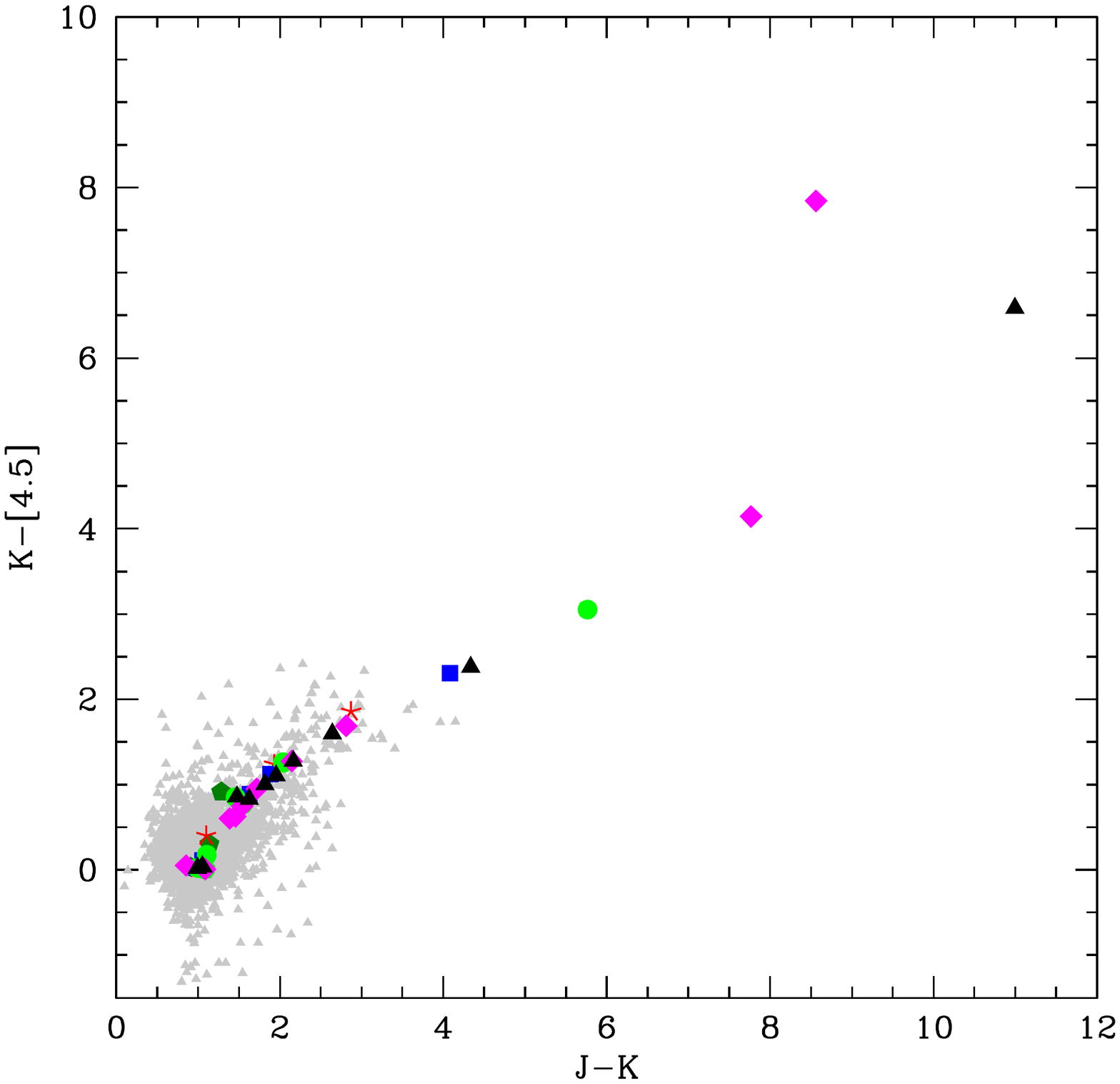}}
\end{minipage}
\vskip-60pt
\caption{The evolution of the luminosity (top, left panel) and of the C/O ratio (top, right) 
of 1-2 M$_{\odot}$ at $Z=10^{-3}$ and 2.5 M$_{\odot}$ at $Z\leq2\times10^{-3}$. The thin, 
horizontal lines mark the location of the TRGB and the boundary between M stars and carbon 
stars. The variation of the (K-[4.5]) colour is shown in the bottom, left panel. We also 
show the excursion of the tracks in the colour-colour (\it{J}-\it{K}) vs (\it{K}-[4.5]) 
plane, overimposed to the data of the IC10 evolved stellar population by \citet{gerbrandt15} 
and \citet{boyer15a}. The adopted reddening is $E(B-V)=1.14$ mag (see section \ref{reddening})}
\label{fmod}
\end{figure*}

The two top panels of Fig.~\ref{fmod} show the variation of the luminosity and of the
C/O ratio during the TP-AGB phase\footnote{We generally refer to AGB to describe the
evolutionary phases that follow the end of core helium burning. We use the terminology
TP-AGB to address the part of the AGB evolution following the first thermal pulse.
The AGB phases preceding the first thermal pulse are called early-AGB (EAGB).} 
of stellar models of metallicity $Z\leq2\times10^{-3}$ and mass in the range 
$1-2.5~M_{\odot}$. In the abscissa we indicate the time, counted since the first
thermal pulse. For clarity reasons we report only the values of the physical quantities 
referring to an evolutionary stage in the middle of each interpulse phase, chosen in 
conjunction with the maximum luminosity. 

The stars reported in Fig.~\ref{fmod} evolve through the two major phases of core H- and 
He-burning with time scales between 400 Myr and 6 Gyr. Fig.~\ref{fmod} shows that the 
overall duration of the TP-AGB phase is in the range 1-2 Myr, and follows a trend that is 
not monotonic with mass. For stars with mass below $\sim 2~M_{\odot}$, the evolution time 
grows with the stellar mass because the higher the mass, the longer the time required 
to eject the external envelope. On the other hand, in the mass domain $M > 2~M_{\odot}$, 
the most relevant factor is that more massive stars evolve to obtain bigger 
cores and larger 
luminosities, which makes the duration of the AGB phase shorter. This behaviour was 
thoroughly described by \citet{ventura16}, where the interested reader can also find a 
discussion regarding the uncertainties of the evolutionary time scales.

In the top, left panel of Fig.~\ref{fmod} we also show the luminosity of the
TRGB ($\sim 2200$ L$_{\odot}$), obtained by following the RGB evolution of low-mass 
models sharing the same chemical 
composition of the AGB models, until the helium flash. Most of the stars evolve at 
luminosities fainter than the TRGB during the evolutionary phases preceding the 
TP-AGB; the only exception is the $\sim 2.5~M_{\odot}$ star, which is brighter than the 
TRGB for a limited part of the EAGB evolution. This is important for the scope of the 
present work because it allows us to use solely data of stars brighter than the TRGB.

A common property of the stars in this mass domain is the achievement of the
carbon-star stage, as a consequence of a series of third-dredge up (TDU) episodes
(see top, right panel of Fig.~\ref{fmod}). Given the relatively small initial oxygen 
content, the C/O$ > 1$ condition is reached after a few TPs, so that the duration of the 
C-star phase is approximately half the overall duration of the TP-AGB evolution. 
The final C/O, in the range $10-50$, increases with the initial mass of the star,
because stars of higher mass experience a higher number of TDU events before the
envelope is entirely lost \citep{ventura14}. 

The amount of carbon accumulated in the external layers of the star, particularly the 
excess of carbon with respect to oxygen, is of extreme importance for the dust production 
in the circumstellar envelope \citep{fg06, lagadec08, groenewegen07, groenewegen09}. 
The increase in the surface carbon favours the
formation of considerable quantities of solid carbon grains, which grow bigger and
bigger as the wind expands from the surface of the star \citep{fg06, flavia15a, 
flavia15b, ventura16}. This is accompanied by the shift towards longer 
wavelengths of the 
stellar spectrum, owing to the reprocessing of the radiation emitted by the central star 
by the dust particles present in the wind. 

Since the SED shifts to the red when the
C/O$ > 1$ condition is reached, the duration of the evolutionary phase during which
the colours are reddest is extremely short; this is because the formation of great
quantities of dust is accompanied by a significant increase in the rate of mass loss,
which provokes a fast consumption of the external mantle, halting the TP-AGB 
evolution \citep{ventura16}. 

The gradual shift to the IR of the SED of the stars is also clear in the bottom, right
panel of Fig.~\ref{fmod}, which shows the evolution of the stars in the colour-colour
({\it J}-{\it K}, {\it K}-[4.5]) plane. For clarity reasons we show only the 
theoretical tracks of $M \leq 2.5~M_{\odot}$ stars in the bottom, right panel of 
Fig.~\ref{fmod}. As stated earlier in this section, the vast majority of the AGB 
population of IC10 is made up of the progeny of these stars; we will return to this
point in the following sections. The sequences of the various low-mass stars cannot be 
distinguished on this
plane, if not for the extension of the track, which changes according to the amount
of dust formed, hence on the amount of carbon accumulated in the surface regions.

Because the duration of the phase with the largest IR emission is only a small fraction of 
the entire AGB life, for most of the time the stars evolve at approximately constant IR 
colours, namely $J-K \sim 1$ and $K-[4.5] \sim 0$. The shift to the red of the SED occurs only in 
the late TP-AGB phases, after a significant enrichment in the surface carbon occurred (see 
the bottom, left panel of Fig.~\ref{fmod}).

In summary, the most important points relevant to our analysis are:

\begin{enumerate}

\item{All the stars of mass above $\sim 1~M_{\odot}$ evolve through phases characterized
by a significant degree of obscuration of the stellar spectrum, with IR colours
up to $J-K \sim 4$ and $K-[4.5] \sim 2.5$, or even larger in same cases. 
The  \it{K}-magnitude spread of sources reaching these colours can be used as an 
indicator of the duration of the star forming phase for ages in the range 
400 Myr - 6 Gyr, because stars of different mass, which formed in different epochs, 
reach the largest degree of obscuration towards the final evolutionary
phases, at different luminosities (see top, left panel of Fig.~\ref{fmod}).
}

\item{All the stars spend most of the AGB evolution with IR colours $J-K \sim 1$ and 
$K-[4.5] \sim 0$. On the CCD this is expected to produce a clump of objects around
the afore mentioned values, which represent the locus of stars with negligible 
amounts} of dust in their circumstellar envelope.

\item{The higher is the initial mass of the star, the more extended to the red is the
evolutionary track. In particular, it is extremely important to check for the presence 
of objects with $J-K > 4$ and $K-[4.5] > 2.5$. While single stars just entered in the
superwind phase or systems surrounded by an optically thick circumbinary disk can evolve
to such red colours, the detection of a significant number of sources with such red
$J-K $ and $K-[4.5]$ would imply 
the presence of a large number of stars of mass above $2~M_{\odot}$, suggesting that the 
epoch around $\sim 500$ Myr was characterized by a high rate of star formation.
The presence of a relatively large number of these stars in the LMC was invoked by 
\citet{flavia14, flavia15a} to explain the detection of stars with a very large degree 
of obscuration, with IR colours $[3.6]-[4.5] \sim 3$. 
}

\end{enumerate}

\section{IC10: the determination of reddening, distance and SFH from IR observations}
\label{calib}
\subsection{Synthetic population}
To study IC10 we use the same method adopted to interpret IR observations of the
LMC \citep{flavia15a}, SMC \citep{flavia15b} and IC 1613 \citep{flavia16}. This approach is 
based on the AGB and RSG models presented and discussed in Section \ref{ATON} and \ref{models}. 
We assume a metal-poor chemistry, with $Z=10^{-3}$, for ages above 1.5 Gyr, which reflects 
into masses below $\sim 2~M_{\odot}$. For stars younger than 200 Myr (i.e. masses above 
$\sim 4~M_{\odot}$) we assume $Z=4\times 10^{-3}$; an intermediate metallicity, 
$Z=2\times 10^{-3}$, is adopted for the stars formed between 200 Myr and 1.5 Gyr ago. 

Based on the AGB+core helium-burning models we generate the synthetic distribution of 
stars in the  different observational planes, to compare with the observations. To do this
we extract randomly a number of stars in the various epochs, according to the values of 
the SFR and of the mass function. For the latter, we assume a Salpeter law, with index 
$x=-1.3$ \citep{salpeter55}. This choice has no effects on the results obtained.

\subsection{The reddening of IC10}
\label{reddening}
IC10 is located in the outskirts of the Local Group, at low Galactic latitude 
(b = -$3.3^{\circ}$); this 
renders the determination of reddening rather uncertain. The values reported by
\citet{demers04} span the range from $E(B-V)=0.40$ mag to $E(B-V)=1.85$ mag. More
recent results in the literature give $E(B-V)=0.78$ mag \citep{sanna08b} and 
$E(B-V)=0.98$ mag \citep{kim09}.

We base our choice on the results shown in the bottom, right panel of Fig.~\ref{fmod} and
discussed in point (ii) of the previous section: the great majority of low mass stars populate 
the region of the CCD centered at $(J-K, K-[4.5])=(1,0)$. We underline that this is 
independent of the dust model adopted. Therefore, we shifted the observed points on the 
CCD, to make the most populated region to be at $(J-K, K-[4.5])=(1,0)$. This method allowed 
us to deduce the reddening, which is found to be $E(B-V)=1.14$ mag, in agreement with the 
analysis by \citet{sakai99}, based on the study of the Cepheids in IC10. This result 
is also consistent with \citet{schlafly11}, who find $E(B-V)=1.13$ mag for the central $5'$ 
of IC10.

\subsection{The distance of IC10}
\label{distance}
A list of published estimates of the distance of IC10, most in the range 
$0.7-0.8$ Mpc, is given in \citet{demers04}. More recent estimates have been provided
by \citet{vacca07, kim09, goncalves12}. In the present analysis we base the determination 
of the distance
on the position in the CMD of the stars with the largest degree of obscuration. These 
objects, as shown in Fig.~\ref{ftracks}, are located in the right side of the plane and
correspond to the late AGB phases of C-stars, reached by stars with initial
mass in the range $1-2.5~M_{\odot}$. As discussed in the previous section, AGB
modelling can provide reliable estimates of their luminosities. These determinations of
the bolometric magnitudes allow us to predict the position of these stars in the CMD,
for various degrees of obscuration, in turn related to the values of the optical depth.
Accordingly, the locus of C-stars will define a diagonal band in the CMD, extending 
from $K-[4.5] \sim 0$ to the red side of the plane. The details of the SFH 
affect the relative distribution of stars across this band, but has practically no
influence on the $K-[4.5]$ vs $[4.5]$ trend. 

To determine the distance of IC10 we shifted vertically the set of the theoretical tracks 
corresponding to $1-2.5~M_{\odot}$ stars (see Fig.~\ref{ftracks}), until a reasonable 
overlapping with the observed sources is achieved. We obtain the best fit of the data 
with a distance of $0.77$ Mpc, in agreement with the studies by \citet{richer01} and 
\citet{sanna08a}.

\begin{figure}
\resizebox{1.\hsize}{!}{\includegraphics{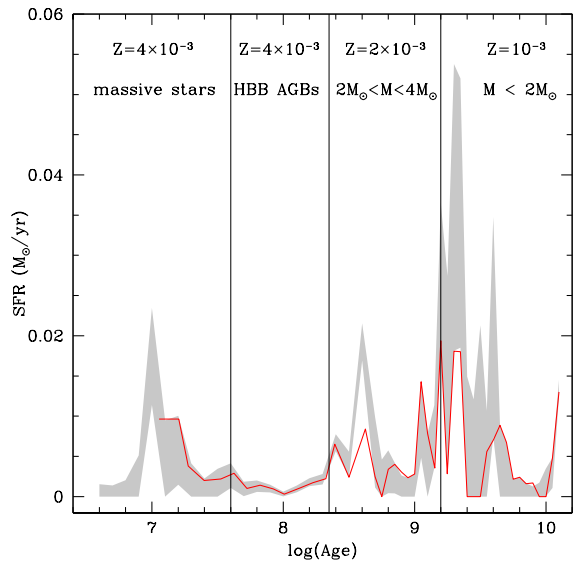}}
\vskip-60pt
\caption{The SFH of IC10 by \citet{weisz14} is shown, as a function of the formation 
epoch of the stars. The grey shaded area delimits the lower and upper limits. The red line 
reports the final choice of the SFH that better reproduce the distribution of the data in 
the colour-magnitude diagrams.}
\label{fsfh}
\end{figure}

\begin{figure*}
\begin{minipage}{0.48\textwidth}
\resizebox{1.\hsize}{!}{\includegraphics{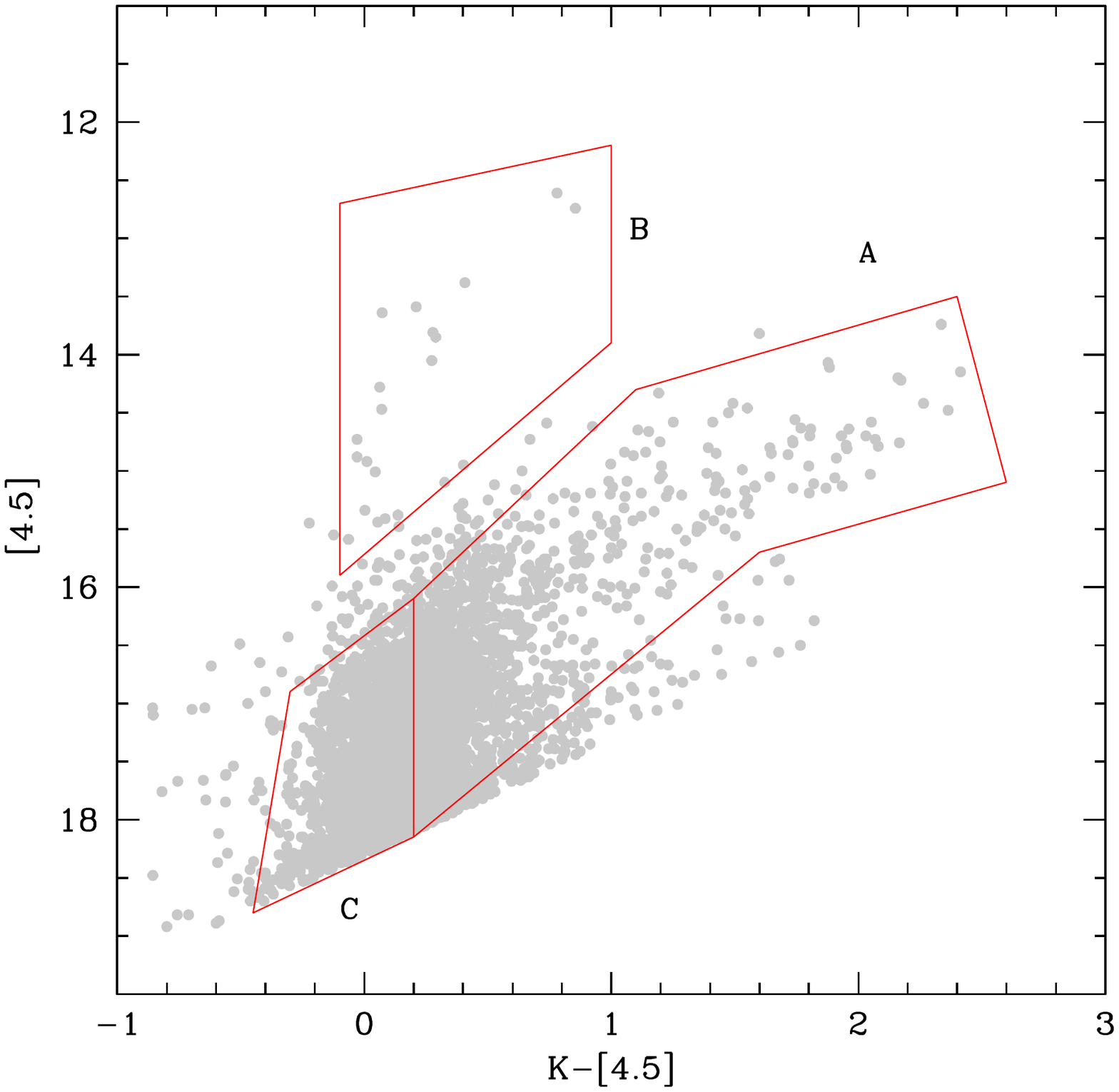}}
\end{minipage}
\begin{minipage}{0.48\textwidth}
\resizebox{1.\hsize}{!}{\includegraphics{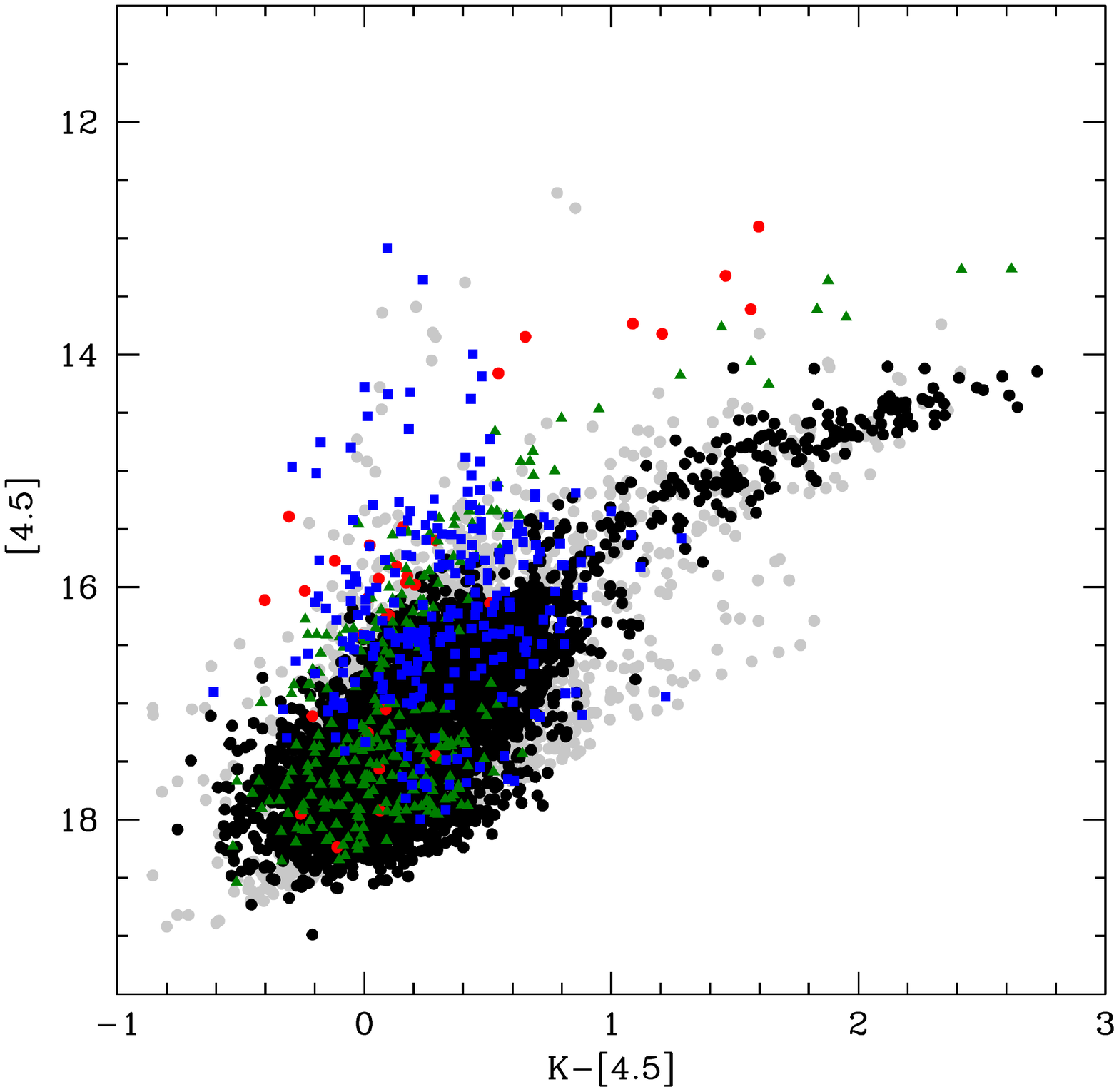}}
\end{minipage}
\vskip-50pt
\caption{Left: The distribution of IC10 stars in the colour-magnitude 
$(K-[4.5],[4.5])$ plane. Right: The results from population synthesis, obtained 
by adopting the recommended SFH by Weisz et al. (2014). The different 
colours indicate stars of metallicity $Z=10^{-3}$ (black points, initial
masses below $2~M_{\odot}$), $Z=2\times 10^{-3}$ (green triangles, descending from
stars of mass $2~M_{\odot} < M < 4~M_{\odot}$), $Z=4\times 10^{-3}$ (red points,
initial mass above $4~M_{\odot}$), core 
helium-burning stars (blue squares, also assumed to have a metallicity
$Z=4\times 10^{-3}$). The observed sample is also shown as light-grey 
points in the right panel.}
\label{fweisz}
\end{figure*}

\subsection{The SFH of IC10}
We adopted the SFH for IC10 published by \citet{weisz14}, shown in Fig.~\ref{fsfh}. 
The grey region delimits the lower and upper limits given by the authors. 
The plane is divided into four regions, according to the 
metallicity and the mass of the stars formed at different times, from the present
day to $\sim 12$ Gyr ago. As a first try, we considered the average SFR between the 
lower and upper limits of Weisz et al. (2013). 

The results are  
shown in Fig.~\ref{fweisz}, where we compare the observed distribution of stars
(left panel), and the synthetic population, reported on the right panel (note that in
the latter panel we aso show the observed stars, with light, grey points). 
In the left panel we indicate three regions of the plane, coded as A, B and C. 
According to our interpretation, region A is populated by stars with a significant 
IR emission, mainly carbon stars, with a small fraction of core helium-burning stars. 
In region B we find bright RSG ($M \geq 12~M_{\odot}$) and massive AGB stars. Region C 
is populated by stars of various 
masses and metallicities, characterized by a negligible IR 
color excess, suggesting scarcity of dust in their surroundings.

\begin{figure*}
\begin{minipage}{0.48\textwidth}
\resizebox{1.\hsize}{!}{\includegraphics{fcmd2obs.pdf}}
\end{minipage}
\begin{minipage}{0.48\textwidth}
\resizebox{1.\hsize}{!}{\includegraphics{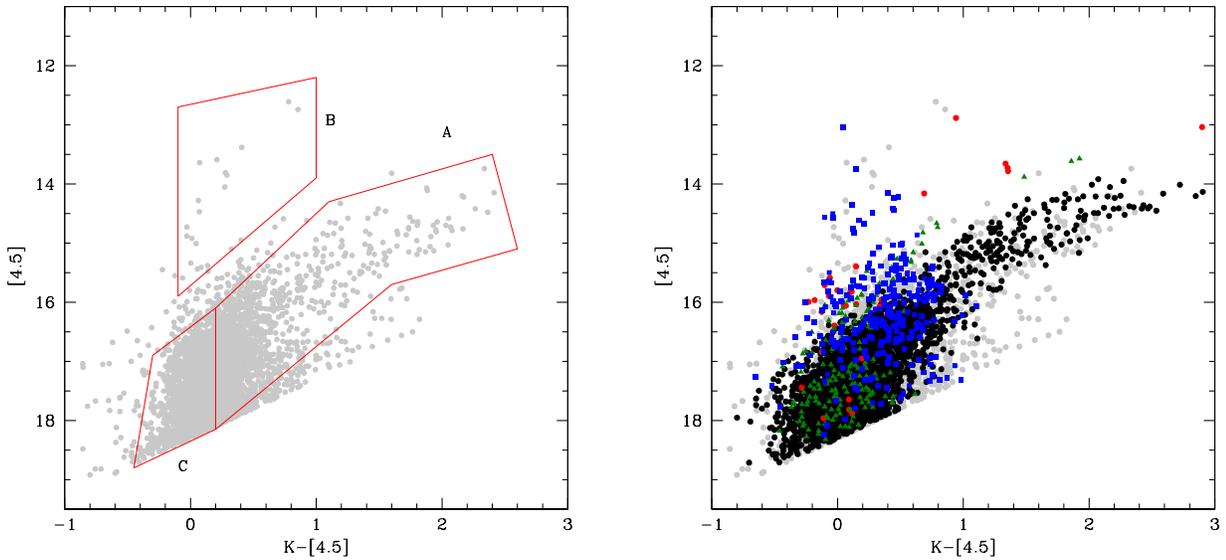}}
\end{minipage}
\vskip-50pt
\caption{The same as Fig.~\ref{fweisz}, with the difference that here the synthetic 
population is obtained by using the SFH represented by the red line in Fig.~\ref{fsfh}. The meaning of
the different colours and symbols is the same as Fig.~\ref{fweisz}.}
\label{fsint}
\end{figure*}

The comparison with the evolutionary 
tracks, presented in Fig.~\ref{ftracks}, shows that 
only stars older than $\sim 500$ Myr evolve into region A of CMD (see Fig. \ref{fsint}). 
This is a relevant information to examine in more details the most remote epochs. In 
particular, the distribution of the tracks (see Fig.~\ref{ftracks}) shows that the width 
of the C-stars locus is associated to a spread in the initial mass of the star, hence in 
the age. A sharply peaked SFH, similar to the one obtained when considering the upper 
limits in Fig.~\ref{fsfh}, would reflect into a narrow distribution of stars in the CMD, 
as shown in Fig.~\ref{fweisz}: we note a strong concentration around the track of 
$\sim 1.3~M_{\odot}$ stars, determined by the peak in the SFH which would have occurred 
$\sim 2$ Gyr ago. On the contrary, we see in the left panel of Fig.~\ref{fweisz}  
a spread in $[4.5]$ slightly in excess of 1 mag, significantly higher than the typical 
photometric error, which is below $0.1$ mag for the stars populating the region of the CMD at 
$K-[4.5] > 1$. This is consistent with a more uniform SFR across the age interval 1-5 Gyr, 
as indicated by the the red line in Fig.~\ref{fsfh}.

We refined the SFH 
relying on the relative fractions of stars in different colours and magnitudes bins and 
on the number counts of stars in the A-B-C zones of CMD. Looking at the comparison between 
the data and the theoretical population, a general agreement is found considering the 
limits suggested for the SFH. Nevertheless, the specific analysis computed in the present 
work allows a more detailed refinement of the SFR, particularly in some age intervals.
The main result of the present analysis is the determination of the SFH for IC10,
represented by the red line in Fig.~\ref{fsfh}. 

The epochs between 300 Myr and 1 Gyr ago are characterized by the formation of stars
with mass in the range $2-3~M_{\odot}$, which reach the C-star stage during their AGB
phase, similarly to their lower-mass counterparts. As discussed in section \ref{models}, 
this class of objects are expected to reach extremely red IR colours (see bottom, left 
panel of Fig.~\ref{fmod}) in the very final AGB phases. The data suggest that only a very 
small number of objects exhibit such a strong IR emission, thus indicating that the SFR 
was not significant in those epochs: our simulations suggest that the SFR must have been 
below $10^{-2}~M_{\odot}$/yr in the 300 Myr - 1 Gyr age interval. This is consistent with 
the prediction by \citet{weisz14}, apart from the higher peak at $\sim$400 Myr found by 
the authors.

In agreement with the recommended values by \citet{weisz14}, we find that the epochs
between 40 Myr and 200 Myr ago were characterized by poor star formation. This is 
the formation epoch of stars of mass in the range $4-8~M_{\odot}$, which are expected to 
achieve copious production of silicates during the HBB phase, reaching a significant
degree of obscuration \citep{ventura14}. A large SFR in those epochs would reflect into 
the presence in CMD of bright, obscured stars. The lack of stars in the region of
CMD included between the tracks of the $2.5~M_{\odot}$ and $7.5~M_{\odot}$ models,
shown in Fig.~\ref{ftracks}, indicate that only a very small number of these stars are 
currently evolving in IC10 (see region B in Fig. \ref{fweisz}).

As stated previously, IC10 is generally believed to have been recently exposed to 
significant star 
formation, which might be still active. It is therefore important to understand whether 
the data set used here can add some information on the star formation activity in recent 
epochs, earlier than $\sim 40$ Myr. We find that an estimate of the rate with which
stellar formation occurred in recent times can be obtained by taking into account the group 
of stars populating the region $0 < K-[4.5] < 0.5$, $[4.5] \sim 16$ in the CMD. We left on 
purpose these sources out of regions A, B and C in Fig.~\ref{fweisz}, 
because, as shown in Fig.~\ref{ftracks}, their position in the plane is consistent both 
with stars of mass in the range $2.5-3~M_{\odot}$ and with RSG stars with masses 
$9-11~M_{\odot}$. In the former case we should expect that significant star formation
occurred 250-600 Myr ago; however, the magnitude spread and the colour extension of the 
locus of carbon stars populating region A indicate that only a limited number of 
$2.5-3~M_{\odot}$ stars are currently evolving in IC10 (see point iii in section 
\ref{models}); therefore, we suggest that these objects are RSG stars. Our simulations 
indicate that the data can be reproduced by assuming significant star formation, with 
rates of the order of $10^{-2}~M_{\odot}/yr$, between 20 and 30 Myr ago, in excellent 
agreement with the values given in \citet{weisz14}.

\section{The evolved stellar population of IC10}
\label{sint}
Based on the choices regarding reddening, distance and star formation history 
discussed in the previous Section, we used a population synthesis approach to 
obtain the expected distribution of the evolved stars in the CMD of IC10.
The results from modelling are shown in the right panel of Fig.~\ref{fsint} 
and compared with the observed points (grey points). We may characterize the
stars observed according to their position on the CMD, which is related to the
mass of the progenitors and the evolutionary status. We divide the stars in the 
groups discussed below.

\begin{figure*}
\begin{minipage}{0.48\textwidth}
\resizebox{1.\hsize}{!}{\includegraphics{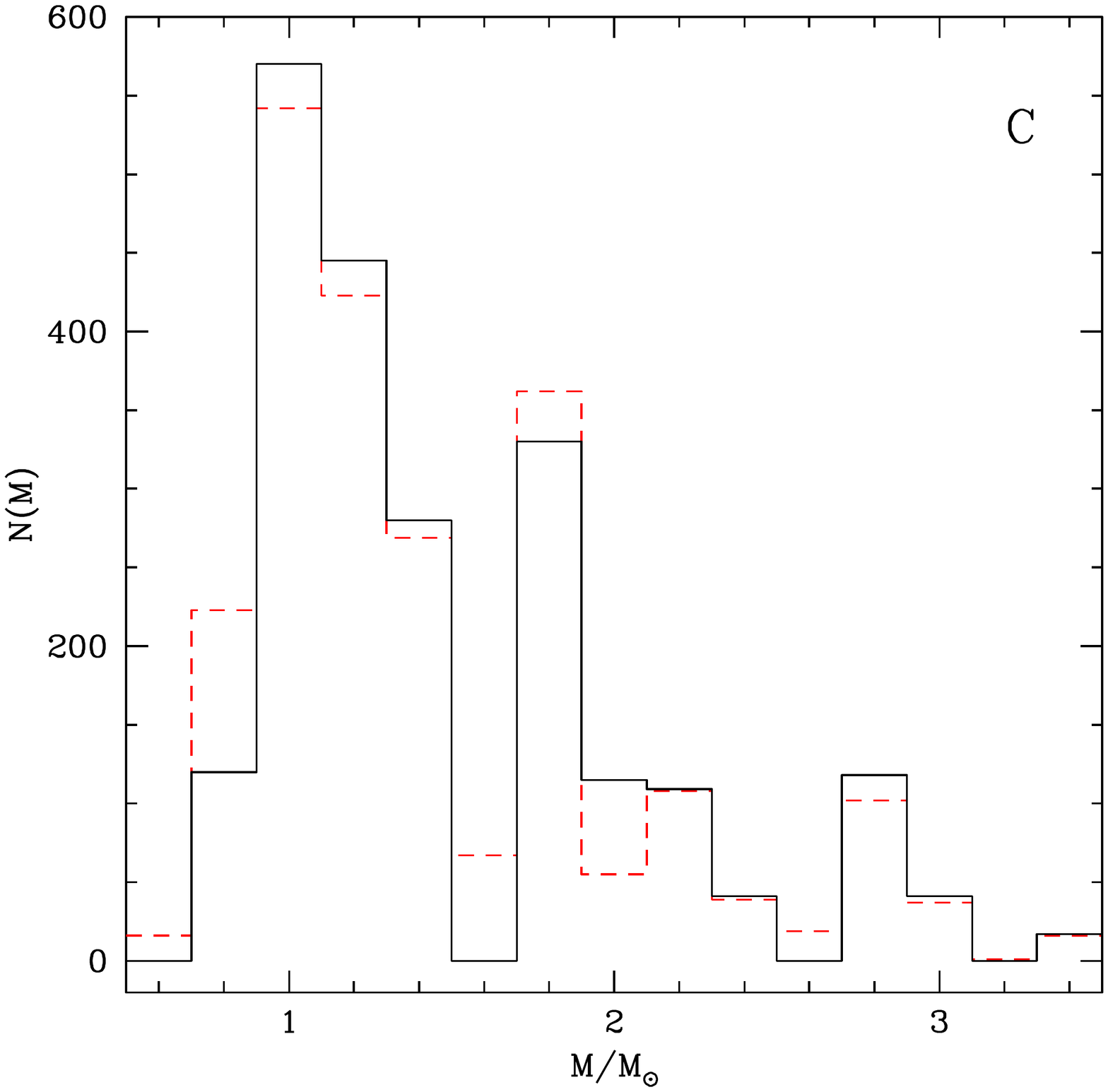}}
\end{minipage}
\begin{minipage}{0.48\textwidth}
\resizebox{1.\hsize}{!}{\includegraphics{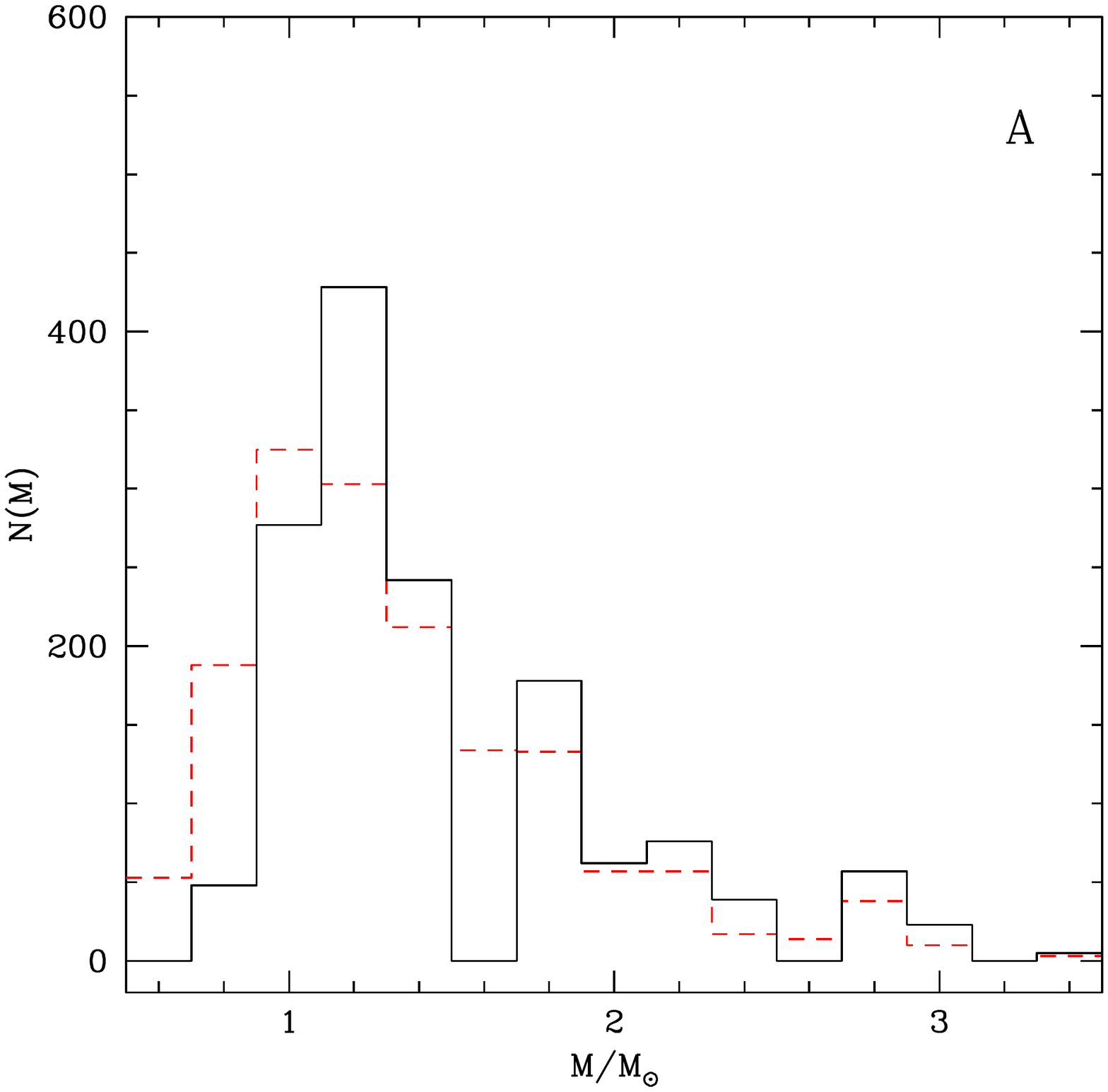}}
\end{minipage}
\vskip-50pt
\caption{The progenitors mass distribution (black, solid lines) and the current mass
(red, dashed) of the stars populating region C (left panel)
and A (right) in Fig.~\ref{fsint}.}
\label{fisto}
\end{figure*}

\subsection{Carbon stars}
Region A in the CMD is predominantly populated by low-mass AGB stars during the C-star 
phase. These sources are the progeny of $1-2.5~M_{\odot}$ stars, formed between $\sim 500$ 
Myr and $6$ Gyr ago. 

The mass histogram of the progenitors of these stars, indicated with a solid line 
in the right panel of Fig.~\ref{fisto}, shows a mass distribution extending from 
$\sim 1~M_{\odot}$ to $\sim 3~M_{\odot}$, 
with a main peak around $\sim 1.2-1.3~M_{\odot}$, a consequence of the maximum in the SFH
that occurred around $\sim 2.5$ Gyr ago (see Fig.~\ref{fsfh}). Secondary peaks in the
mass distribution of the progenitors of these stars are found at $\sim 1.75~M_{\odot}$
and $\sim 3~M_{\odot}$, due to the star formation activity that took place,
respectively, 1Gyr and 300 Myr ago.

The stars populating region A with $1 < K-[4.5] < 2.5$ are experiencing the
last TPs, before entering the PNe phase. The surface C/O ratio is in the range
$5 <$C/O$< 10$ (see Fig.~\ref{fmod}). The current mass of these sources spans the 
interval $0.6-0.7~M_{\odot}$. Although the interpretation of these observations is
partly affected by the photometric uncertainties, we predict brighter stars
are of higher initial mass (c.f Fig.~\ref{ftracks}), because higher mass stars
are brighter during these late AGB phases. A small fraction of stars ($<1\%$ of 
the entire sample) populate the area below region A, with $1 < K-[4.5] < 2$. From 
a deeper inspection based on different colour-magnitude planes, they are probably 
residual unresolved foreground sources, background galaxies and young stellar 
objects \citep{boyer11}. The lack of photometric measurements with filters at 
wavelengths longer than 4.5 $\mu $m (e.g. ($J$-[8]) affects the possibility 
to further reduce this contamination.

Region A is also populated by RSG stars of mass in the range
$6-9~M_{\odot}$ (see the shaded region in Fig.~\ref{ftracks}). The expected number of 
these objects within the box delimiting region A partly depends on the assumptions 
regarding the optical depth. With the present hypothesis on $\tau_{10}$, discussed in 
Section \ref{ATON}, we find that the percentage of core helium-burning stars in region A is 
$\sim 5\%$. The choice of smaller optical depths would lead to a smaller fraction of
core helium-burning stars in this zone of the CMD. 

Independently of the $\tau_{10}$ choice, 
the dominant population of stars in region A is made up of low-mass, carbon stars.
Although the expected number of core helium-burning stars is small in all cases, exploring this
possibility is important to understand whether this zone of the CMD harbours
carbon stars solely, or whether a few obscured O-rich stars surrounded by silicate-type dust 
are also present.

Region A harbours $\sim 40\%$ of the overall evolved population of IC10; 
this result, in agreement with the observations, is consistent with the results 
shown in Fig.~\ref{fmod} and discussed in section \ref{models}, indicating that low mass, 
metal-poor stars evolve as carbon stars for almost half of the AGB phase.

\subsection{The bright, young stars}
\label{bright}
The stars populating region B have luminosities above $2\times 10^4~L_{\odot}$,
which rules out the possibility that they are carbon stars (see the top, left
panel of Fig.~\ref{fmod}). Their magnitudes are compatible either with oxygen-rich 
AGB stars, descending from $M \geq 4~M_{\odot}$ progenitors, currently experiencing HBB,
or RSG stars, with mass in the range $10-20~M_{\odot}$ (see the shaded region in 
Fig.~\ref{ftracks}). Selecting between these two possibilities is not obvious. However, 
we believe that most of the stars evolving into region B, with the exception of the two 
brightest objects, are RSG stars, based on two arguments:

\begin{itemize}
\item{
The stars observed in region B define an approximately vertical sequence, 
indicating a scarce degree of obscuration; as shown in Fig.~\ref{ftracks},
this is at odds with the expectations concerning the evolution of massive AGB stars,
which produce significant amounts of silicates during the AGB phase \citep{ventura14}, 
thus evolving at
redder colours. A possible solution would be that these stars are metal poor, but this
would imply that no metal enrichment occurred in IC10, which seems unlikely.
}

\item{
As discussed previously, IC10 is known to harbour a relatively young stellar population, 
as a consequence of high star formation in recent epochs: therefore, the stars in region B 
are more likely the signature of this recent star forming activity. 
}
\end{itemize}

\subsection{Scarcely obscured stars}
Region C in Fig.~\ref{fsint} harbours stars with a very small degree of obscuration, 
indicating scarcity of dust in their circumstellar envelope. The progenitor mass 
distribution, shown in the left panel of Fig.~\ref{fisto}, is dominated by low-mass stars, 
with mass $M \leq 1.5~M_{\odot}$. 

Compared to the mass distribution of stars evolving to region A,
we note a higher fraction of $M < 1~M_{\odot}$ objects. This difference occurs because
stars of mass below $\sim 1~M_{\odot}$ are not expected to reach the C-star stage, 
thus they evolve through the AGB phase as oxygen-rich objects, almost dust-free.
Note that stars of mass $M \leq 1~M_{\odot}$ evolve brighter than the TRGB for only
a part of the AGB phase (see the $1~M_{\odot}$ model track in Fig.~\ref{ftracks});
the mass distribution of region C would be more peaked towards the lowest masses
otherwise.

We note in the left panel of Fig. \ref{fisto} that the secondary peaks around 
$\sim 2~M_{\odot}$ and $\sim 3~M_{\odot}$ are higher than in region A (right panel). 
This is because the stars within this range of mass are expected to evolve into region
C of the CMD not only in the initial part of the TP-AGB phase, during which little
dust is produced, but also for a significant portion of the early AGB phase, also
characterized by negligible formation of dust. This is different from their lower mass
counterparts, which are expected to evolve brighter than the TRGB (thus within region C)
only during the TP-AGB phase.

Region C also involves a modest fraction (below $\sim 5\%$) of core helum-burning stars
of mass $6~M_{\odot} < M < 9~M_{\odot}$, with ages younger than $100$ Myr.

\subsection{Scarcely obscured, bright stars: RSG or low-mass objects?}
The analysis of the stars observed in the regions of CMD between zones B and
C is particularly tricky, because these sources could either descend from stars of mass 
in the range $2.5-3~M_{\odot}$, which have just entered the C-star phase (see the track of 
the $2.5~M_{\odot}$ model in Fig.~\ref{ftracks}), or by RSG stars with mass
$9~M_{\odot} < M < 11~M_{\odot}$ (see the shaded region in Fig.~\ref{ftracks}).

In the former case this region of the plane would be populated by stars of mass just below 
the threshold required to activate HBB, formed between 300 Myr and 1 Gyr ago; this possibility 
would require a higher SFR in these epochs, compared to the choice shown in Fig.~\ref{fsfh}. 
However, we are much more favourable to consider contamination from RSG stars, 
because: i) the presence of such class of stars would require the existence of stars 
brighter than the group of sources populating region A, which is indeed not observed; 
ii) stars of mass $2-3~M_{\odot}$ reach very large degrees of obscuration, 
which would reflect into a higher fraction of stars at IR colours (see Fig. \ref{ftracks}) 
with respect to the observational evidence.

\begin{figure*}
\begin{minipage}{0.48\textwidth}
\resizebox{1.\hsize}{!}{\includegraphics{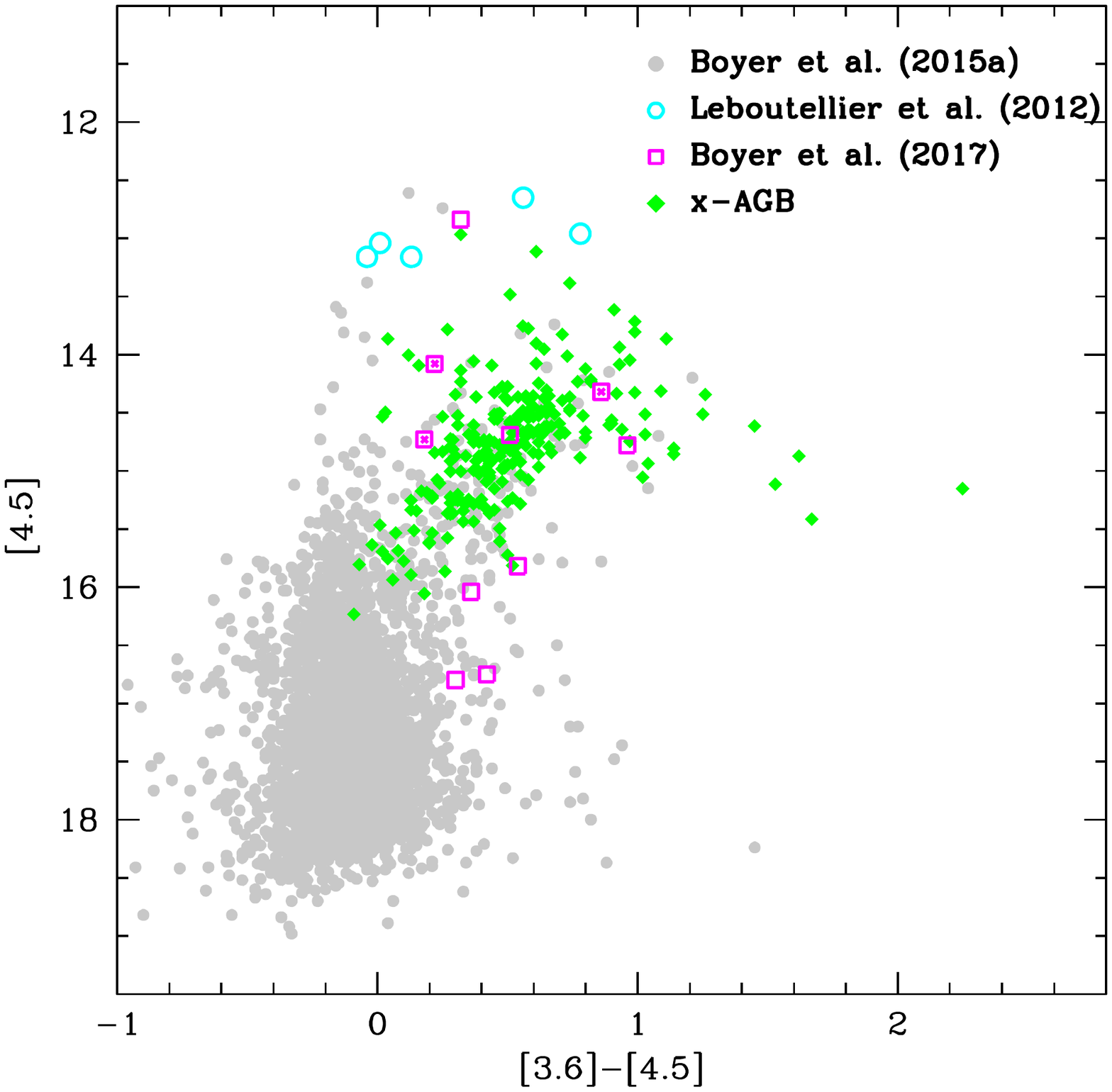}}
\end{minipage}
\begin{minipage}{0.48\textwidth}
\resizebox{1.\hsize}{!}{\includegraphics{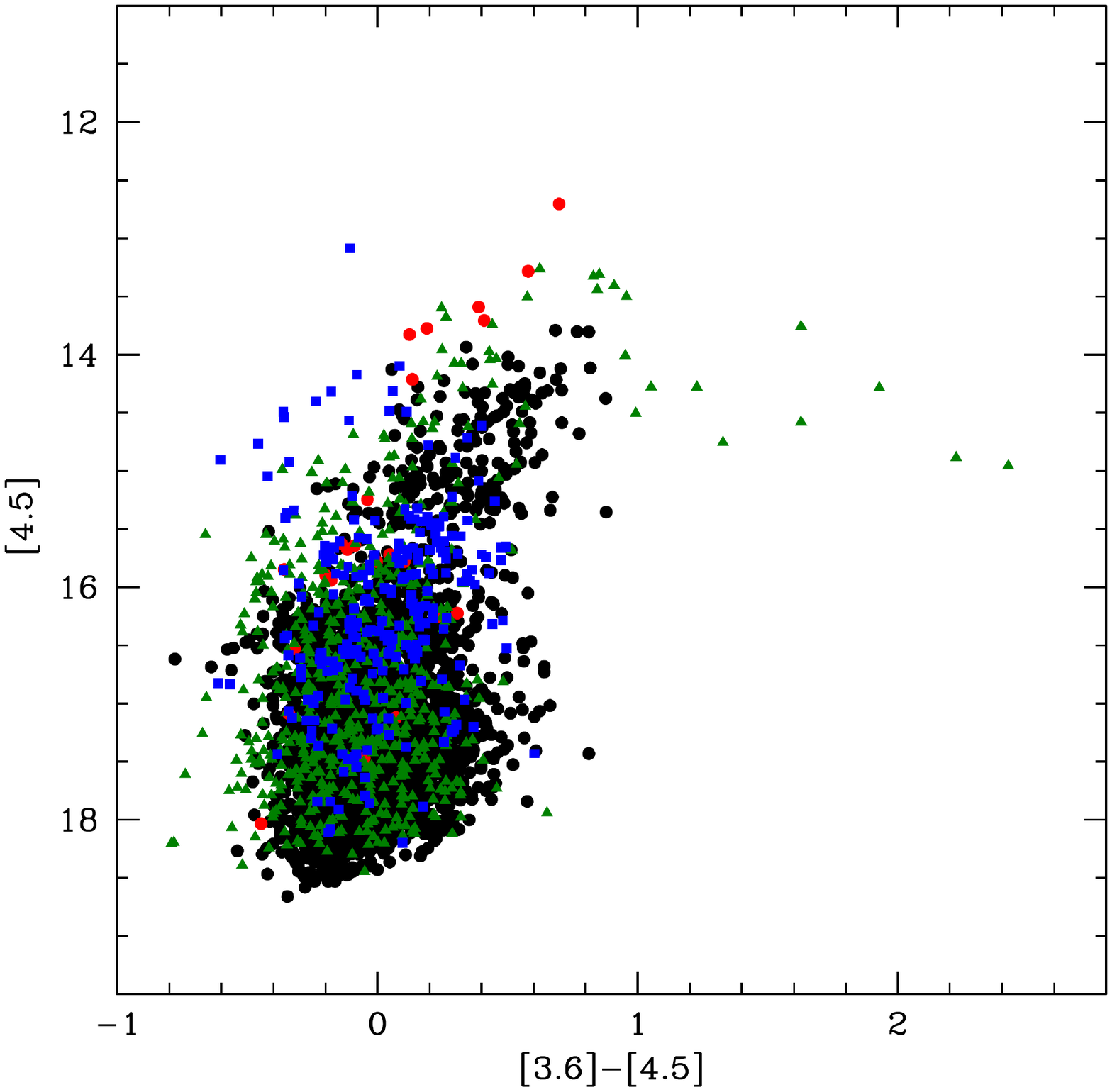}}
\end{minipage}
\vskip-50pt
\caption{Left: The distribution on the colour-magnitude $([3.6]-[45],[4.5])$ plane of the IC10
stars from \citet{boyer15a}, indicated with grey points. Cyan circles indicate the bright, 
O-rich stars by \citet{lebo12}, whereas the obscured stars by \citet{boyer17}, classified 
as O-rich, are indicated with magenta squares (crosses inside the square mean variable
sources). The x-AGB stars by \citet{boyer15b} are represented with light green diamonds.
Right: The results from synthetic modelling, with the same symbols used in
Fig.~\ref{fweisz} and \ref{fsint}.}
\label{fspitzer}
\end{figure*}

\section{An analysis of obscured stars in the \emph{Spitzer} colour-magnitude plane}
\label{dustystars}

To study in a more complete way the population of evolved stars in IC10
we study their distribution on the colour-magnitude ($[3.6]-[4.5],[4.5]$) plane 
(hereafter CMDSP). We attempt an analysis similar to the one based on the 
combined near-IR and \emph{Spitzer} data, done in the previous section on the CMD. 
This complementary study allows us to consider the RSG, M-type and xAGB candidates 
identified by \citet{lebo12} and \citet{boyer15b,boyer17}, as described in 
Section \ref{sample}.

The left panel of Fig.~\ref{fspitzer} shows the distribution in the CMDSP of the stars 
in the sample by \citet{boyer15a}, the x-AGB stars by \citet{boyer15b}, the stars
studied by \citet{lebo12} (only the 5 stars for which the [4.5] data are available are 
shown) and the dusty, O-rich sources by \citet{boyer17} (we restrict 
our attention to the 10 sources classified as M-type with a high confidence level).
The right panel of Fig.~\ref{fspitzer} reports the results obtained with our synthetic 
approach, described in details in the previous sections, used to produce the results
shown in the right panel of Fig.~\ref{fsint}. 

\begin{figure}
\resizebox{1.\hsize}{!}{\includegraphics{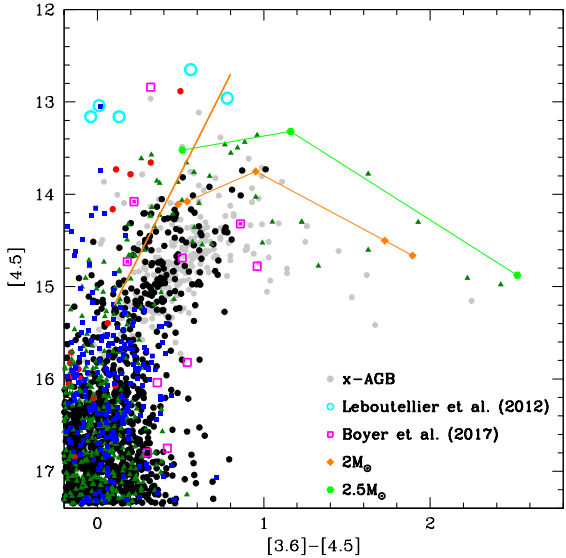}}
\vskip-60pt
\caption{The distribution of IC10 x-AGB stars by \citet{boyer15b} (grey points), the stars
from \citet{lebo12} (cyan circles) and \citet{boyer17} (magenta squares) on the 
colour-magnitude $([3.6]-[45],[4.5])$ plane. Results from synthetic modelling are also
shown, with the same symbols of Fig.~\ref{fweisz} and \ref{fsint}. The evolutionary 
tracks of stars of initial mass $2~M_{\odot}$ (orange) and $2.5~M_{\odot}$ (light green)
are shown.}
\label{freddened}
\end{figure}

Fig.~\ref{freddened} show the observations and the results from population synthesis at
the same time; to better characterize the obscured stars we focus on a narrower region
of the CMDSP(grey points).

We distinguish a sequence of obscured 
objects, extending to $[3.6]-[4.5] \sim 1$, which we interpret mainly as carbon stars.
Furthermore, we note a sequence of bright sources, with $[3.6]-[4.5] < 0.3$, which
corresponds to the stars populating region B in Fig.~\ref{fsint}: based on the arguments
given in Section \ref{bright}, we believe that the majority of these sources are RSG
stars, with a few AGB stars experiencing HBB.

The diagonal, orange line in Fig.~\ref{freddened} can be used to separate carbon
stars, expected to be on the right of the line, and their O-rich counterparts, which
populate the region on the left. RSG stars can be found both on the left and on the right 
side of the line. According to our interpretation the majority ($>90\%$) of the sources 
observed by \citet{boyer15b} are carbon stars, descending from low-mass progenitors, of 
mass in the range $1-2.5~M_{\odot}$. 

The stars exhibiting the largest IR emission are of particular
interest in the present context, because IC10 harbours the largest number of
objects with $[3.6]-[4.5] > 1$, among all the galaxies in the DUSTiNGS survey \citep{boyer15b}.
For these sources, which are surrounded by large amounts of dust,
the colour provides and indication of the mass of the progenitor, because
as far as C-stars are concerned, stars of higher mass reach higher degree of
obscuration in the late AGB phases, owing to a larger accumulation of carbon in 
the surface regions \citep{ventura16}. This is partly shown in the right panel of 
Fig.~\ref{fspitzer}, where we report the tracks in the CMDSP of stars of mass 
$2, 2.5~M_{\odot}$, during the final part of the AGB evolution; lower mass stars do not 
reach such red colours.

Based on these arguments, we conclude that the stars with the largest degree of
obscuration in the \citet{boyer15b} sample are the progeny of $2-2.5~M_{\odot}$ stars,
formed between 500 Myr and 1 Gyr ago. They are evolving through the final AGB
phases, after the loss of a significant fraction of their envelope, such that their
current mass is in the range $0.6-0.9~M_{\odot}$. The surface regions of these stars
are enriched in carbon, owing to the action of repeated TDU events: the surface
mass fractions and the C/O ratios are, respectively, $X(C) \sim 0.01-0.015$ and
C/O$ \sim 3-5$. Their large IR colours are determined by the great amounts of dust 
formed, mainly under the form of solid carbon grains of $0.15-0.25 \mu$m size.

The variable sources by \citet{boyer15b} on the left side of the diagonal line are
massive AGB stars, currently experiencing HBB. A more detailed characterization of the
progenitors is not straightforward in this case, because all the stars of initial mass
$M \geq 4~M_{\odot}$ evolve to those regions of the CMDSP.

The M-type, dusty stars identified by \citet{boyer17}, indicated with magenta, open squares, 
are distributed across the CMDSP (see right panel of Fig.~\ref{fspitzer} and
Fig.~\ref{freddened}), spanning a wide range of colours ($0.3<[3.6]-[4.5]<1$) 
and magnitudes ($12.5<[4.5]<17$). The five sources with IR colours $0.3<[3.6]-[4.5]<0.6$
populate a region of the plane distant from the zone covered by the tracks of massive
AGB stars. While their colours could be reproduced by invoking a degree of obscuration
of AGB stars much larger than the model predictions, their magnitudes are too faint to be
considered as stars undergoing HBB. Based on their position on the CMDSP, we
may consider the possibility that these sources descend from stars of mass in the
range $7-10~M_{\odot}$, currently evolving through the core He-burning phase. This
possibility requires a degree of obscuration unusually large for this class of
objects, of the order of $\tau_{10} \sim 0.2$. Further possibilities to explain these
objects are mixed-chemistry sources or binary systems with induced mass loss.

\begin{figure*}
\begin{minipage}{0.48\textwidth}
\resizebox{1.\hsize}{!}{\includegraphics{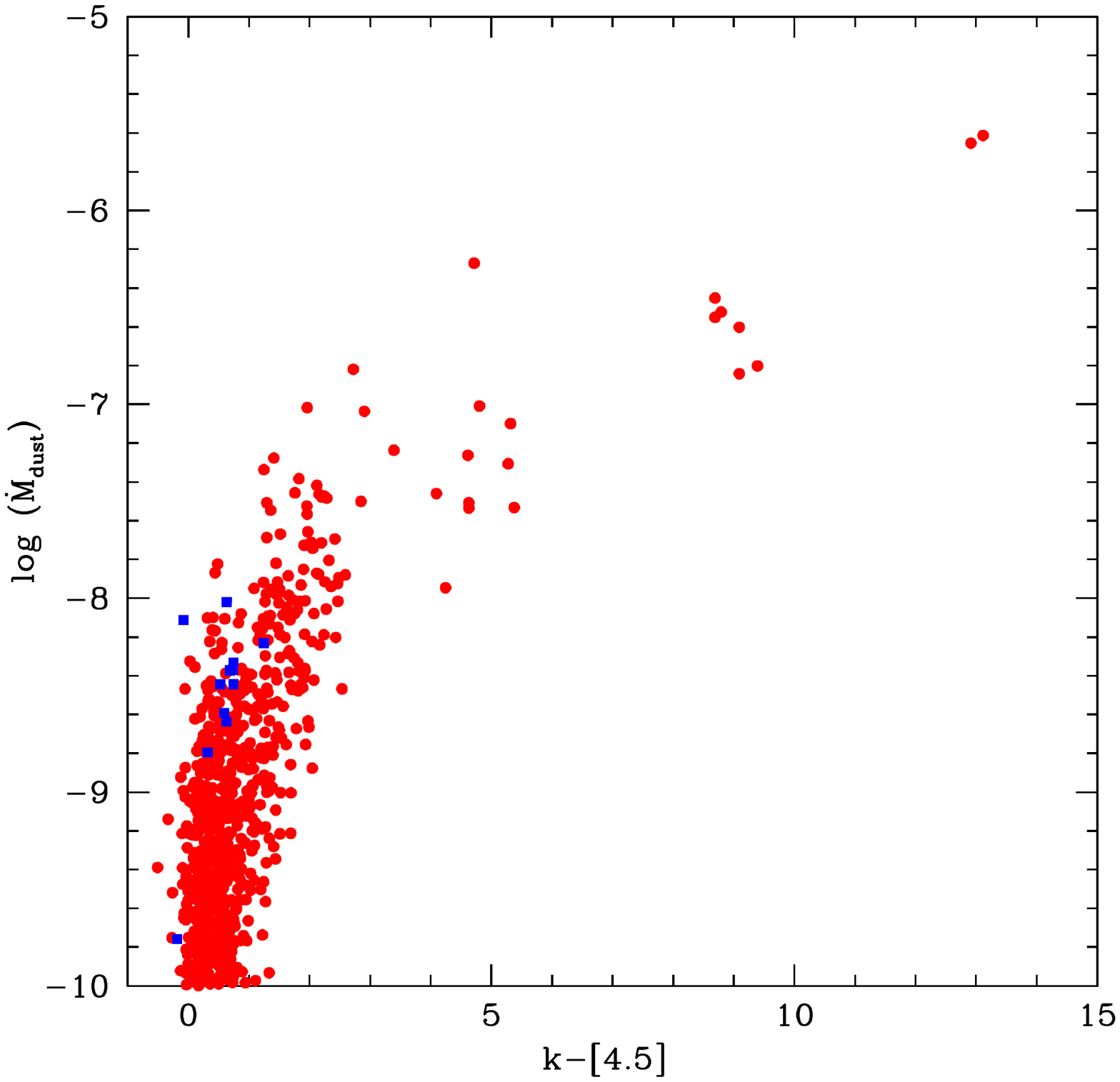}}
\end{minipage}
\begin{minipage}{0.48\textwidth}
\resizebox{1.\hsize}{!}{\includegraphics{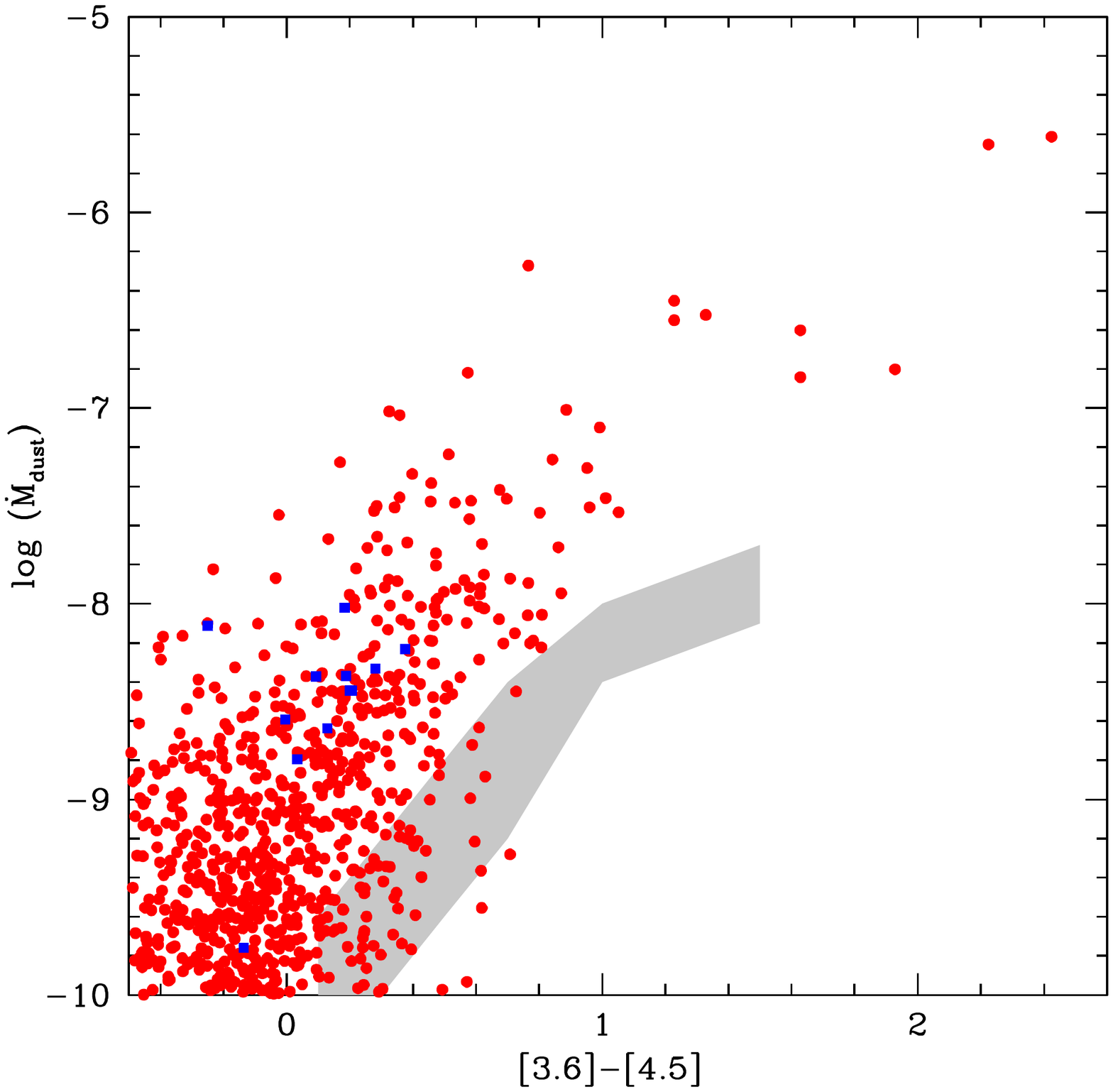}}
\end{minipage}
\vskip-50pt
\caption{The dust mass loss rate of the stars in our synthetic population, as a 
function of the $K-[4.5]$ (left panel) and $[3.6]-[4.5]$ (right) colours. Red points 
indicate carbonaceous dust, whereas blue squares indicate silicate and alumina dust 
production. The grey, shaded region in the right panel indicate the relation for SMC 
AGB stars, given in \citet{srinivasan16}.}
\label{fdpr}
\end{figure*}

The three stars on the left of the diagonal line 
in the CMDSP are AGB stars evolving through the HBB phase. For the two faintest objects 
(\#112431 and \#118138) this possibility is supported by the DUSTiNGS classification 
as x-AGB variables. The luminosities of the two sources is
compatible with any mass higher than $\sim 4~M_{\odot}$. \#105975 is the
brightest among this sub-sample: based on the position on the CMDSP, we conclude that
it is a massive AGB star, descending from a $M > 5~M_{\odot}$ progenitor. Note that
\citet{boyer17} claimed a progenitor mass above $\sim 8~M_{\odot}$ for this object.
The dissimilarity in this interpretation is due to the higher luminosities achieved
by the massive AGB models used in the present analysis, determined by the use of the
full spectrum of turbulence model for turbulent convection \citep{cm91}, which favours 
stronger HBB conditions (hence
higher luminosities) for $M \geq 4~M_{\odot}$ stars compared to the models
based on the traditional mixing length theory \citep{vd05, ventura15}.

The interpretation of the two stars from \citet{boyer17} with the largest IR colours shown in
Fig.~\ref{freddened}, namely \#121876 and \#120247, is more cumbersome, given that their 
position rules out the possibility that they are massive AGB stars. This conclusion is 
pretty independent of the details of the dust formation modelling: our tests show that 
an artificial increase in the optical depth would lead to small values of $[4.5]$, at 
least one magnitude brighter than observed. Their colours and magnitudes can be reproduced 
by RSG models of mass around $\sim 8-10~M_{\odot}$, if an optical depth of 
$\tau_{10} \sim 1$ is assumed. This very large degree of obscuration might be 
associated to a strong mass loss occurring during the RSG evolution. For the star \#121876 
this explanation is at odds with the variability detected. We leave this problem open.

Regarding the five stars analysed by \citet{lebo12}, for which the combined [3.6] 
and [4.5] data are available, we see in Fig.~\ref{freddened} that they populate a
large luminosity region in the CMDSP, with their $[3.6]-[4.5]$ colors spanning almost
1 mag. The bluer three sources, i.e. \#4, \#6, \#14, populate the region of the 
CMDSP where stars of mass around $20~M_{\odot}$ evolve during the RSG phase. This 
understanding is in agreement with the conclusions drawn by \citet{lebo12} regarding 
the nature of these objects. Conversely, the IR fluxes of the redder two stars,
\#8 and \#12, are hardly compatible with an RSG origin. Their colours and luminosities 
are consistent with massive AGB stars undergoing HBB, with progenitor masses 
$M \geq 6~M_{\odot}$. This conclusion is also supported by the $K-[4.5]$ colours of 
these two objects, of the order of $K-[4.5] \sim 1.5$, much more consistent with a 
massive AGB than with an RSG origin (see the track of the $7.5~M_{\odot}$ model in 
Fig.\ref{ftracks}).

\section{Dust production in IC10}
The analysis done in the previous sections outlined that IC10 harbours a numerous
group of carbon stars, with various degrees of obscuration. The increase in
the IR emission in these objects is triggered by the formation of solid
carbon grains that grow bigger in size. In this context, we may neglect the formation
of SiC grains, owing to the low metallicity involved, hence a low Si/C ratio
\citep{sloan12, ventura14}.

According to our interpretation, the majority of the stars populating region A in the CMD,
shown in Fig.~\ref{fsint} as black points, are surrounded by carbon dust grains, whose 
size reaches $\sim 0.15 \mu$m in the reddest objects, populating the region of the plane
with $K-[4.5] \sim 2.5$. The optical depths of these stars ranges 
from $\tau_{10} \sim 0.2$ to  $\tau_{10} \sim 1$. While most of the evolved stellar population
of IC10 is shown in Fig.~\ref{fsint}, the results presented in the previous section 
and shown in Fig.~\ref{fspitzer} indicate the presence of a few stars with an 
extremely large degree of obscuration, which populate the region of CMDSP at 
$1.5 < [3.6]-[4.5] < 2.5$. Some of these sources are not present in the
\citet{gerbrandt15} sample, owing to their low fluxes in the {\it K} band.
These stars, classified as extreme AGB by \citet{boyer15b}, 
are characterized by optical depths of the order of $\tau_{10} \sim 4$ and are
surrounded by carbon grains of $\sim 0.25 \mu$m size.

The current dust mass loss rates for the AGB stars in the synthetic population of IC10 
are shown as a function of the IR colours in Fig.~\ref{fdpr}. Note that only the stars with rates above
$10^{-10}$ M$_{\odot}$/yr are shown. Thus the dust mass loss rates by low-mass stars
during the initial AGB phases, before becoming carbon stars, are excluded from this
figure.
 
According to our interpretation, stars populating region A in Fig.~\ref{fsint} 
eject dust into the interstellar medium with rates in the range 
$10^{-8} \rm M_{\odot}/\rm yr < \dot{M}_{\rm d} < 3 \times 10^{-7} \rm M_{\odot}/yr$. As stated
previously, this dust is essentially under the form of solid carbon particles.
The stars with the reddest colours are characterized by higher rates, up
to $\dot{M}_{\rm d} \sim 2 \times 10^{-6} \rm M_{\odot}/yr$.

Fig.~\ref{fdpr} (right panel) includes a comparison to the SMC carbon-stars locus
from \citet{srinivasan16}, showing significant differences
in the $[3.6]-[4.5] > 0.5$ domain. Our results for the dust mass loss rates are
5-10 times higher than those given in \citet{srinivasan16}; more important, the slope
of the $\dot{M}_{\rm d}$ vs $[3.6]-[4.5]$ trend is larger in this study.
These dissimilarities mainly reflect the differences in the outflow velocities: while 
\citet{srinivasan16} assume $\rm v_{\rm out} = 10$ km/s, we find that the speed of the
winds changes with the degree of obscuration, ranging from $\rm v_{\rm out} \sim 5-10$ km/s
for the carbon stars with the lowest IR emission, to $\rm v_{\rm out} \sim 40-50$ km/s
for the most obscured stars. A further though less relevant factor is $\Psi$, the 
dust to gas ratio: while \citet{srinivasan16} assume $\Psi = 0.005$ for carbon stars,
in the present study we find that $\Psi$ ranges from = $0.002$ to $0.01$. 

The discussion in the previous sections showed that massive AGB stars are present in
very modest numbers in IC10. These sources are represented as blue squares in
Fig.~\ref{fdpr}. The overall dust production rate (DPR) by this class of objects is
therefore negligible when compared to the carbon dust budget. Additional dust, but
in negligible quantities, is produced by low-mass stars during the AGB phases previous 
to the achievement of the carbon star stage. Regarding RSG stars, we do not have
a straight recipe to find out the rate with which they form dust. However, we expect a
minor contribution, because these sources trace an almost vertical trend both
in the CMD and CMDSP, thus suggesting that only small quantities of dust are present
in their surroundings. This is consistent with the conclusions by \citet{lebo12}.

Based on the results shown in Fig.~\ref{fdpr} we estimate an overall DPR by the evolved 
stellar population of IC10 of $7 \times 10^{-6} \rm M_{\odot}$/yr.
This dust is mainly composed of solid carbon grains. Silicate particles provide
a very small contribution, of the order of $10^{-8} \rm M_{\odot}$/yr.

\begin{figure}
\resizebox{1.\hsize}{!}{\includegraphics{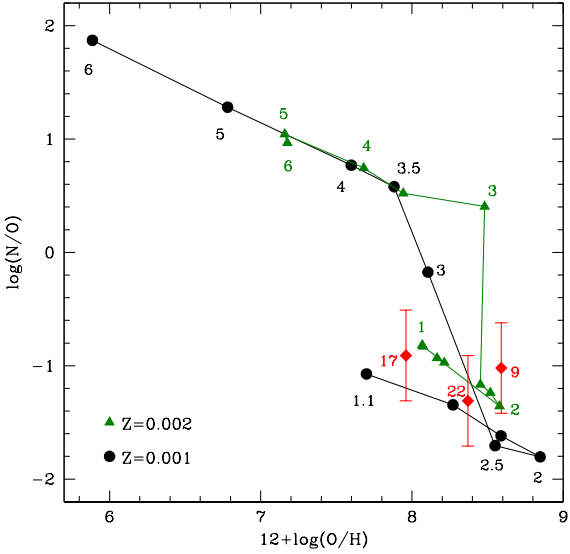}}
\vskip-60pt
\caption{Red diamonds indicate the chemical composition of the 3 PNe in IC10 studied by 
\citet{magrini09}, for which both the O and N measurements are available. Black points
and green triangles points the final chemistry of AGB models of mass in
the range $1-6~M_{\odot}$ and metallicity, respectively, $Z=10^{-3}$ and  
$Z=2\times 10^{-3}$.}
\label{fpne}
\end{figure}

\section{The PNe population of IC10}
\citet{magrini09} discussed the further possibility offered by the analysis of
the PNe population to reconstruct the SFH of IC10. In their
work the authors present spectroscopic observations of twelve PNe in IC10,
nine out of which were spectroscopically confirmed.

In order to compare our predictions and understanding of the evolved population 
nowadays evolving in IC10 with the results from \citet{magrini09}, we rely
on the three PNe for which both the N and O abundances are available, namely
PN9, PN17 and PN22; given the lack of the carbon mass fractions, the 
contemporary knowledge of the abundances of these two species is the minimum 
requirement to achieve any interpretation.

Fig.~\ref{fpne} shows the data by \citet{magrini09} in the oxygen vs $N/O$ plane,
compared to our models, for stars of different initial mass and metallicities
$Z=1,2 \times 10^{-3}$. The trend with mass of the theoretical locii follows
a counterclockwise pattern, originated by the relative importance of TDU and HBB 
during the AGB phase for stars of different mass \citep{ventura15b}. 

In agreement with the conclusions by \citet{magrini09} we rule out the possibility
that the three PNe descend from massive progenitors, because the measured $N/O$'s
are not compatible with any contamination from HBB, which would favour a 
large increase in the N content, hence in the $N/O$ ratio. This may be consistent
with the discussion in Section \ref{sint}, according to which we expect only
a very few massive AGB stars to evolve in IC10. Anyway, the PNe sample is far from being 
complete or representative, therefore we can not draw any conclusion from this evidence.

PN17 is the one with the smallest oxygen among the PNe in the sample. We 
tentatively suggest that it descends from a low-metallicity
progenitor, of initial mass around $\sim 1.1~\rm M_{\odot}$. The chemistry of 
PN22 and PN9 is compatible with progenitors of slightly higher mass, of the
order of $\sim 1.3-1.5~\rm M_{\odot}$. This understanding is further supported by the 
observed fluxes from these objects. To this aim, we extended the evolutionary tracks 
until the beginning of the WD cooling (Ventura et al. in prep.) and verified that the post-AGB luminosities
of $1.1~\rm M_{\odot}$ and $1.5~\rm M_{\odot}$ stars differ by $\delta \log \rm L = 0.3$.
This is in agreement with the fact that the relative fluxes observed at $\lambda = 5007~\rm{\AA}$ given by \citet{magrini09} are $\log \rm (F_{PN9}/F_{PN17})=0.27$ and $\log \rm (F_{PN22}/F_{PN17})=0.21$.


 \section{Conclusions}
We study the conspicuous evolved stellar population of the star-burst galaxy IC10, with the aim of
characterizing the individual sources observed in terms of mass, age and chemical
composition and to reconstruct the SFH of the galaxy. We use stellar models evolved through the 
AGB phase, which also include the description of the dust formation mechanism.
This step is crucial to interpret the SED and the colours
of these objects, because the dust in their circumstellar envelopes reprocesses the
radiation emitted by the central star to IR wavelengths. 

The analysis is based on the combination of IR and mid-IR fluxes of a sample of stars, 
obtained by cross-correlating different catalogues. The distribution of the stars in
the ($J-K$, $K-[4.5]$) and ($K-[4.5]$, $[4.5]$) planes allows the evaluation of the reddening and 
distance of the galaxy: we obtain $E(B-V)=1.85$ mag and d$=0.77$ Mpc, within the range
of values reported in the literature. 

The analysis of the overall distribution of stars in the ($K-[4.5]$, $[4.5]$) plane
indicates that $\sim40\%$ of the IC10 stars brighter than the TRGB are carbon stars,
descending from $1-2.5~M_{\odot}$ progenitors. The majority of these objects descend from
$1.1-1.3~\rm M_{\odot}$ stars, formed during the major epoch of star formation, which
occurred around 2.5 Gyr ago. The scarcity of stars in the region of the plane
covered by oxygen-rich, massive AGB stars, experiencing HBB, indicates low star
formation between 40 Myr and 200 Myr ago. The presence of bright stars in the observed
distribution can be reproduced by invoking significant star formation, of the order
of $10^{-2}~\rm M_{\odot}$/yr, in recent times ($<40$ Myr); this is consistent with the fact that IC10
is considered as a starburst galaxy.

The stars with the largest IR emission, characterized by extremely low NIR fluxes,
are only present in the \emph{Spitzer} sample. Though few in number, accounting for the 
presence of these objects is important for a correct determination of the current DPR 
in IC10. The comparison with the models suggests that these extremely red sources are 
carbon stars, descending from $2-2.5~\rm M_{\odot}$ stars, and are evolving through the very final 
AGB phases. Their large degree of obscuration is due to the presence of significant 
quantities of carbon dust in their winds, with grains of $0.2-0.25 \mu$m size. These
stars provide the largest contribution to the dust nowadays ejected into the interstellar
medium of this galaxy, with DPR above $10^{-7}~\rm M_{\odot}$/yr.
We estimate that the overall DPR of IC10, largely dominated by carbon stars, 
is $7\times 10^{-6}~\rm M_{\odot}$/yr.

The present work further confirms that the IR study of the evolved population of galaxies 
is a promising tool to analyse the stellar content of the host system, to reconstruct the
star formation history and to provide a determination of the reddening and distance.
This is a welcome result under the light of the upcoming launch of the \emph{JWST} that will 
significantly enlarge the volume of dwarf galaxies where bright and evolved stellar 
populations will be accessible.

\section*{Acknowledgments}
The authors are indebted to the anonymous referee, for the careful reading of the paper
and for the several suggestions and comments, that helped improving significantly
the quality and the clarity of the manuscript.
FDA and DAGH acknowledge support provided by the Spanish Ministry of
Economy and Competitiveness (MINECO) under grant AYA-2017-88254-P.

\end{document}